\documentclass{mn2e}
\usepackage{epsfig,rotating}

\usepackage{mnras_cite}

\newcommand{\asca}{{\it ASCA}}

\newcommand{\sax}{{\it BeppoSAX}}

\newcommand{\rxte}{{\it RXTE}}

\newcommand{\exosat}{{\it EXOSAT}}
\newcommand{\etal}{{et al.}}
\newcommand{\lum}{\thinspace\hbox{$\hbox{erg}\thinspace\hbox{s}^{-1}$}}

\newcommand{\lta}{\hbox{${_<\atop{\sim}}$}}
\newcommand{\fl}{\thinspace\hbox{$\hbox{erg}\thinspace\hbox{cm}^{-2}\thinspace\hbox{s}^{-1}$}}

\newcommand{\gta}{\hbox{${_>\atop{\sim}}$}}
\newcommand{\msun}{\thinspace\hbox{$M_{\odot}$}}
\newcommand{\rsun}{\thinspace\hbox{$R_{\odot}$}}

\title
[Simultaneous X-ray/optical observations of GX\,9+9]
{Simultaneous X-ray/optical observations of GX\,9+9 (4U\,1728--16)}

\author[A. K. H. Kong \etal]
{A. K. H. Kong$^{1,6}$\thanks{Email: akong@space.mit.edu}, 
P. A. Charles$^{2,5}$, L. Homer$^{3,6}$, E. Kuulkers$^{4}$ and D. 
O'Donoghue$^{5}$ \\
$^1$ Kavli Institute for Astrophysics and Space Research, Massachusetts 
Institute of Technology, \\
77 Massachusetts Avenue, Cambridge, MA 02139, U.S.A.\\
$^2$ Department of Physics and Astronomy, University of Southampton, Southampton SO17 1BJ\\
$^3$ Astronomy Department, University of Washington, Box 351580, Seattle, WA 98195-1580, U.S.A.\\
$^4$ ISOC, ESA/ESAC, Urb. Villafranca del Castillo, P.O. Box 50727, 28080 
Madrid, Spain \\
$^5$ South African Astronomical Observatory, PO Box 9, Observatory 7935, 
Cape Town, South Africa\\
$^6$ Department of Astrophysics, University of Oxford, Keble Road,
Oxford OX1 3RH
}
\date{Accepted. Received.}
\begin{document}
\maketitle

\begin{abstract}
We report on the results of the first simultaneous X-ray (\rxte) and
optical (SAAO) observations of the luminous low mass X-ray binary
(LMXB) GX\,9+9 in 1999 August. The high-speed optical photometry
revealed an orbital period of 4.1958~hr and confirmed previous
observations, but with greater precision. No X-ray modulation was
found at the orbital period. On shorter timescales, a
possible 1.4-hr variability was found in the optical light curves which 
might be related to the mHz quasi-periodic oscillations seen in other 
LMXBs. We do not find any significant X-ray/optical correlation in the light curves.
In X-rays, the colour-colour diagram and hardness-intensity diagram
indicate that the source shows characteristics of an atoll source in
the upper banana state, with a correlation between intensity and
spectral hardness. Time-resolved X-ray spectroscopy suggests that
two-component spectral models give a reasonable fit to the X-ray
emission. Such models consist of a blackbody component which can be
interpreted as the emission from an optically thick accretion disc or
an optically thick boundary layer, and a hard Comptonized component
for an extended corona.
\end{abstract}


\section{Introduction}

GX\,9+9 (4U\,1728--16) is a persistent bright galactic bulge X-ray source, with a $4.19\pm
0.02$~hr orbital period that was reported by Hertz and Wood (1988) from
{\it HEAO-1} data. The variations were roughly sinusoidal with an
amplitude of 3.8\%. However, there was clearly substantial scatter (4.6\% residual error) about this modulation. The source showed a  
rich variety of X-ray variability on timescales from minutes to months and 
correlated X-ray spectra and fluxes (\pcite{mason76}; \pcite{parsignault78}).
Observations with {\it EXOSAT} show a spectrum and spectral variations very
similar to GX\,3+1, GX\,13+1 and GX\,9+1 (\pcite{schulz89}) and so GX\,9+9 was classified as a high-luminosity atoll source (Hasinger \& van der Klis 1989). The X-ray spectra
obtained by {\it EXOSAT} and {\it Einstein} are very similar and can be
represented by a blackbody of 1.5~keV and a thermal bremsstrahlung component of
$\sim$~8~keV (\pcite{christian97}). An \asca\ observation
revealed a similar kind of spectrum with a blackbody of $\sim$~1.3~keV plus a cutoff
power law with photon index of $\sim$~1.2 and cutoff energy of $\sim$~6~keV
(\pcite{church00}). Line emission was also found from previous observations
({\it Einstein}: 0.77, 0.82 and 1.06~keV; \pcite{christian97}; {\it ROSAT}:
0.65
and 0.97~keV; \pcite{schulz99}) but a recent \asca\ observation found none
(\pcite{church00}). 
An \rxte\ observation of GX\,9+9 was carried out in 1996
but no kHz QPOs were found (Wijnands, van der Klis \& van Paradijs 1998). There is no report of X-ray bursts from this source. 

The optical counterpart ($V\sim$~17) is a `Sco\,X-1-like' blue star with He
II $\lambda$4686 and C III/N III $\lambda$4640 emission lines
(\pcite{davidsen76}; \pcite{charles77}).
The only temporal study in the optical performed to date is that of 
Schaefer (1990) in which a $\sim$~20\% modulation was present on the X-ray
periodicity. There is no radio detection for GX\,9+9 according to the recent Australia Telescope Compact Array observations at 4.8 and 8.64~GHz \cite{berendsen00}.

There have been no simultaneous optical/X-ray observations of this source so
far. At $V=17$ it is ideal for studying outer disc variability
simultaneously with \rxte\ observations of the X-ray variability. In
this paper, we report the results of the {\it
first} simultaneous X-ray/optical observations of GX\,9+9. An observation log
for both X-ray and optical observations is presented in
\S\ref{sect:obs}, while the temporal analysis of the data is in
\S\ref{sect:time}. In \S\ref{sect:spec}, time-resolved X-ray spectroscopy
is presented and an archival study of X-ray modulation is
reported in \S5.
The implications of our new data will be
discussed in \S\ref{sect:9+9discussion}. 

\section{Observations}
\label{sect:obs}

\subsection{SAAO Observations}

All optical observations were carried out with the UCT-CCD fast photometer \cite{odonoghue95} at the Cassegrain focus of the 1.9-m telescope at the South African Astronomical Observatory (SAAO), Sutherland from 1999 August 11 to 14. The observing conditions were good with typical seeing of
$\sim$~1--2 arcsec and there was $\sim$~6~hr observation time each night. In
general, the seeing during the first two nights was quite stable while the
image quality of the remaining two nights was relatively poor due to the
variable seeing and high wind conditions. In order to achieve high time
resolution, no filter
was employed (i.e. white light) and a $5\times5$ on-chip
prebinning was applied to maximise the time resolution.
The timing resolution of all the data is 2~s. In Table~\ref{tab:9+9log}, an
observing log is presented together with the X-ray
observations. Debiasing and flat-fielding were performed with standard
IRAF routines. Since the star fields are crowded, differential
photometry was obtained via PSF fitting with IRAF. The light curves of
all four nights are shown in Fig.~\ref{fig:lc1} to \ref{fig:lc4} and
have typical error bars on individual 2~s observations of less than 2\%.

Apart from the high-speed photometry, we also obtained a set of $B$
and $V$ images on 1999 August 15. The exposure times on GX\,9+9 were 
60~s in the $B$-band and 40~s in the $V$-band. We calibrated the
magnitudes using a photometric standard, giving $B=16.96\pm0.05$ and $V=16.79\pm0.05$.

\subsection{\rxte\ Observations}

GX\,9+9 was observed with the Proportional Counter Array (PCA) instrument on board the \rxte\ \cite{bradt93} between 1999 August 11 and August 14 (see
Table~\ref{tab:9+9log}). The PCA consists of five identical
Proportional Counter Units (PCUs) sensitive to X-rays with an energy
range of 2--60~keV and a total effective area of
$\sim$~6500~cm$^2$. The PCA spectral resolution at 6~keV is
approximately 18 per cent, and the maximum timing resolution available
is 1\,$\mu$s. Three PCUs were on during our observations.
For most of the time, \rxte\ observed the source
simultaneously with the optical telescope at SAAO. The typical on-source time for each night is about 10--14~ks (see Table~\ref{tab:9+9log}). Apart from the standard
configuration of the PCUs (Standard1 and Standard2 modes which are always available), we also employed event modes with
maximum timing resolution of 125\,$\mu$s and 64 spectral energy
channels in order to maximise the timing and spectral resolution. All
X-ray spectra presented here have been corrected for background and
dead-time with FTOOLS v4.2 and Epoch 4 PCA background model. 

\begin{table}
\caption{A Journal of the X-ray/optical observations of GX\,9+9.} 
\label{tab:9+9log}
\begin{center}
\begin{tabular}{c c c c c} \hline
Observatory&Detector& Date&\multicolumn{2}{c}{Time (UT)}\\
&                                          &         &Start &End\\
\hline
SAAO& 1.9m +             &1999 August 11&      17:26&23:08\\
&     UCT CCD                         &1999 August 12&      17:15&23:08\\
&                              &1999 August 13&      17:16&23:00\\
&                              &1999 August 14&      17:45&23:05\\
\rxte\ & PCA		       &1999 August 11&      17:20&00:06\\
&                              &1999 August 12&      17:28&00:03\\
&                              &1999 August 13&      17:27&22:41\\
&                              &1999 August 14&      17:26&22:40\\
\hline
\end{tabular}
\end{center}
\end{table}   

\section{Timing Analysis}
\label{sect:time}

\subsection{Optical}

\begin{figure*}
\begin{center}
\psfig{file=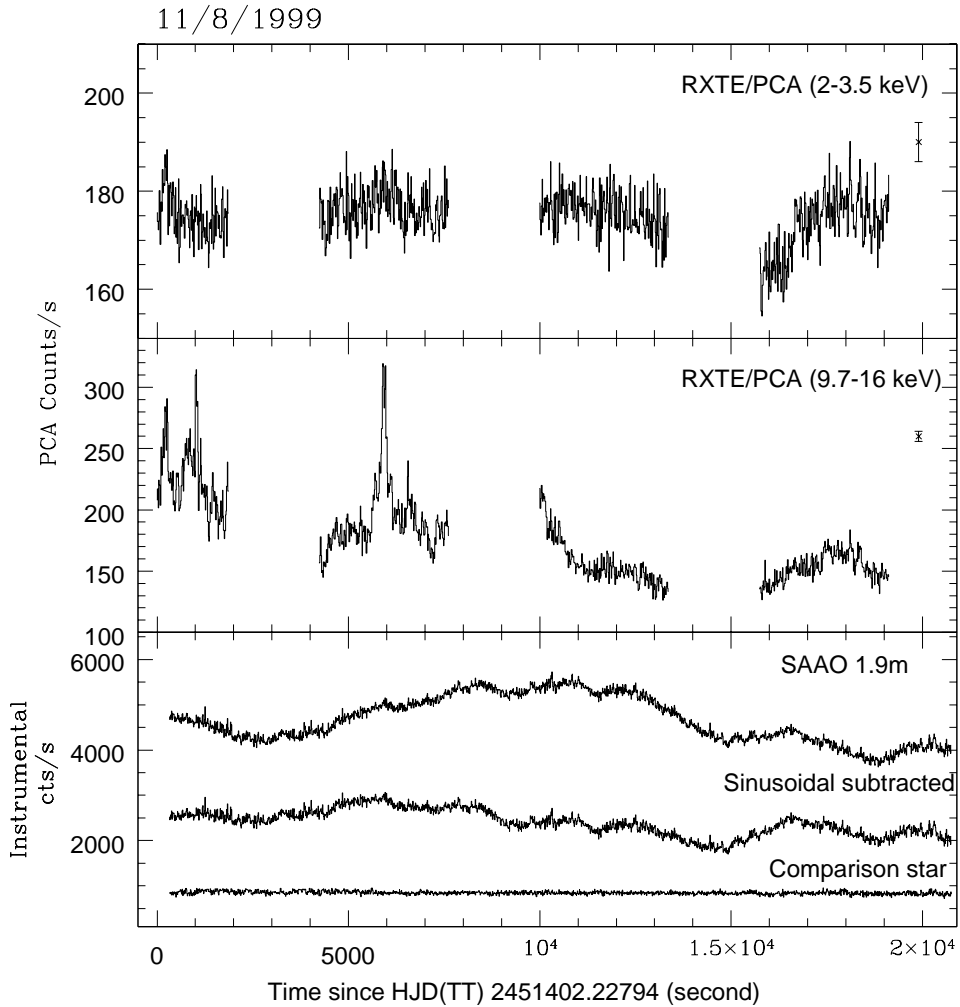,height=14cm}
\caption{Simultaneous X-ray and optical light curves of GX\,9+9 on 11/8/99. The X-ray
light curve was divided into two energy bands: 2--3.5~keV (upper panel)  and 9.7--16~keV (middle panel); typical error bars for the X-ray data are also shown.
The timing resolution is 16~s for X-ray data and 2~s for optical data. For clarity, the optical light curve (lower panel) has been rebinned to 10~s. Note that a 4.2-hr orbital modulation is obvious in the optical light curve.  Also shown underneath the source is a light curve with the 4.2-hr sinusoidal modulation subtracted. Optical light curves of a comparison star are also plotted.} 
\label{fig:lc1}
\end{center}
\end{figure*}

\begin{figure*}
\begin{center}
\psfig{file=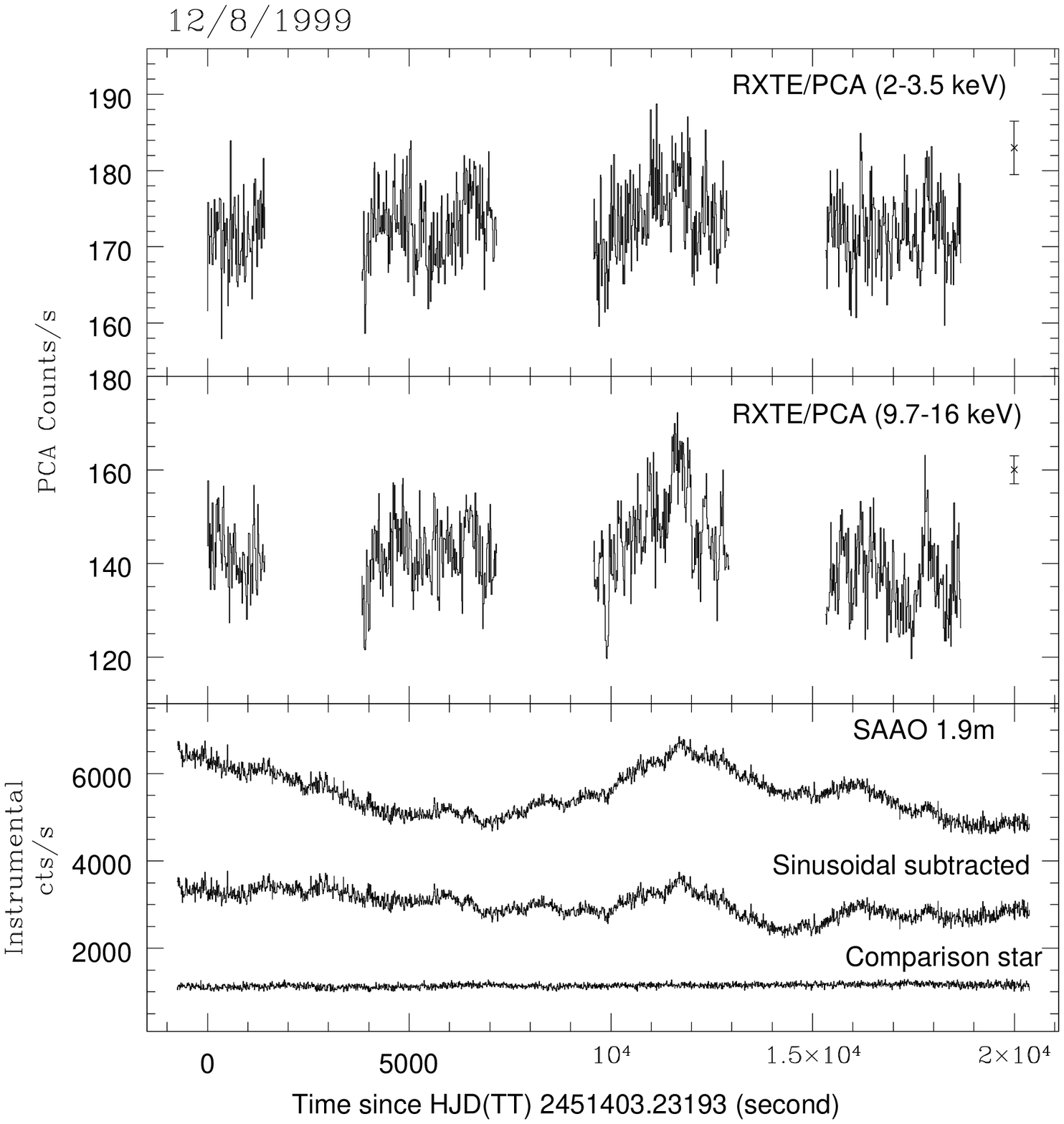,height=14cm}                       
\caption{Same as Fig.~\ref{fig:lc1}, but for 12/8/99.}
\label{fig:lc2}
\end{center}
\end{figure*}

\begin{figure*}
\begin{center}
\psfig{file=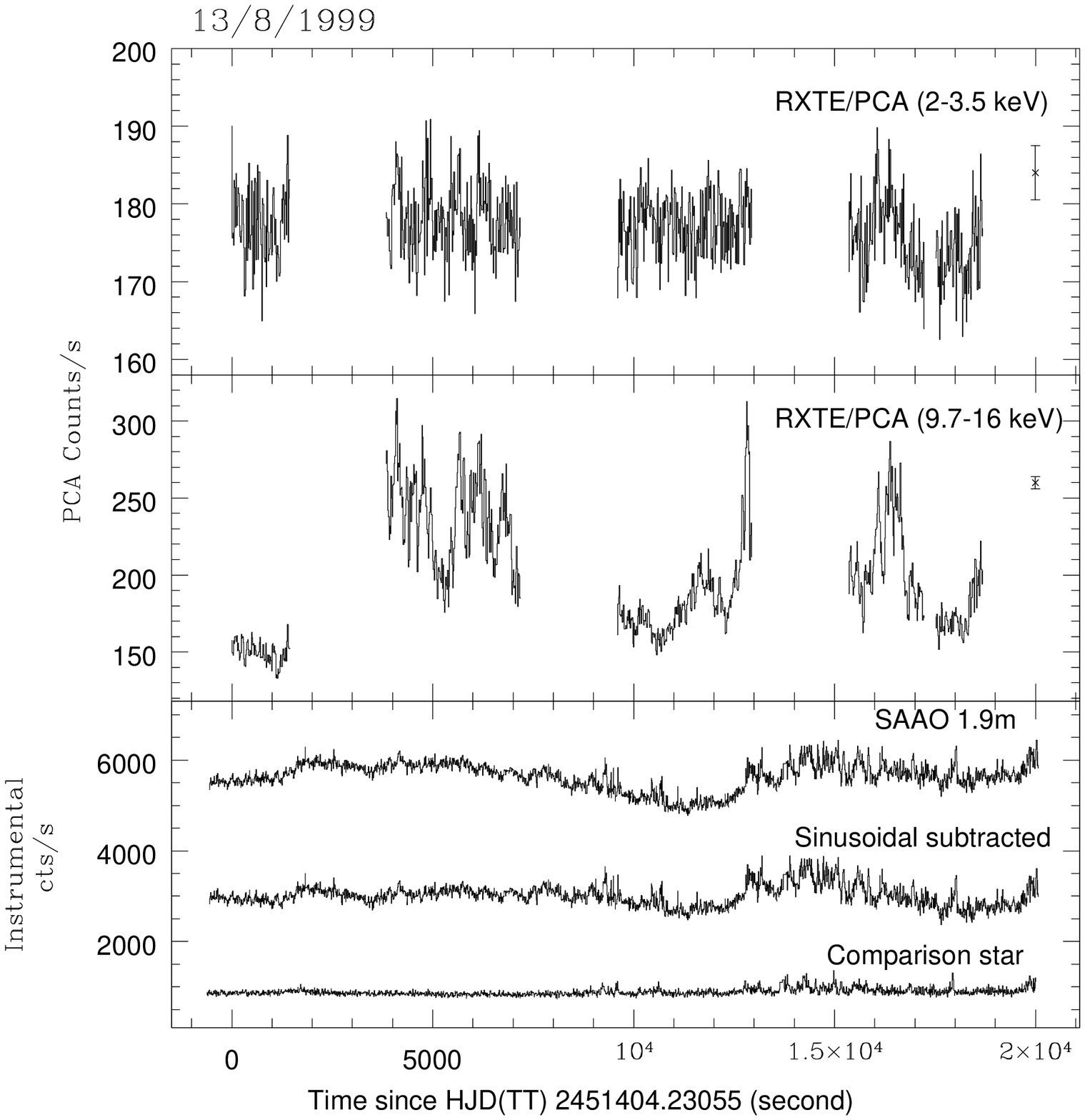,height=14cm}                       
\caption{Same as Fig.~\ref{fig:lc1}, but for 13/8/99.}
\label{fig:lc3}
\end{center}
\end{figure*}

\begin{figure*}
\begin{center}                       
\psfig{file=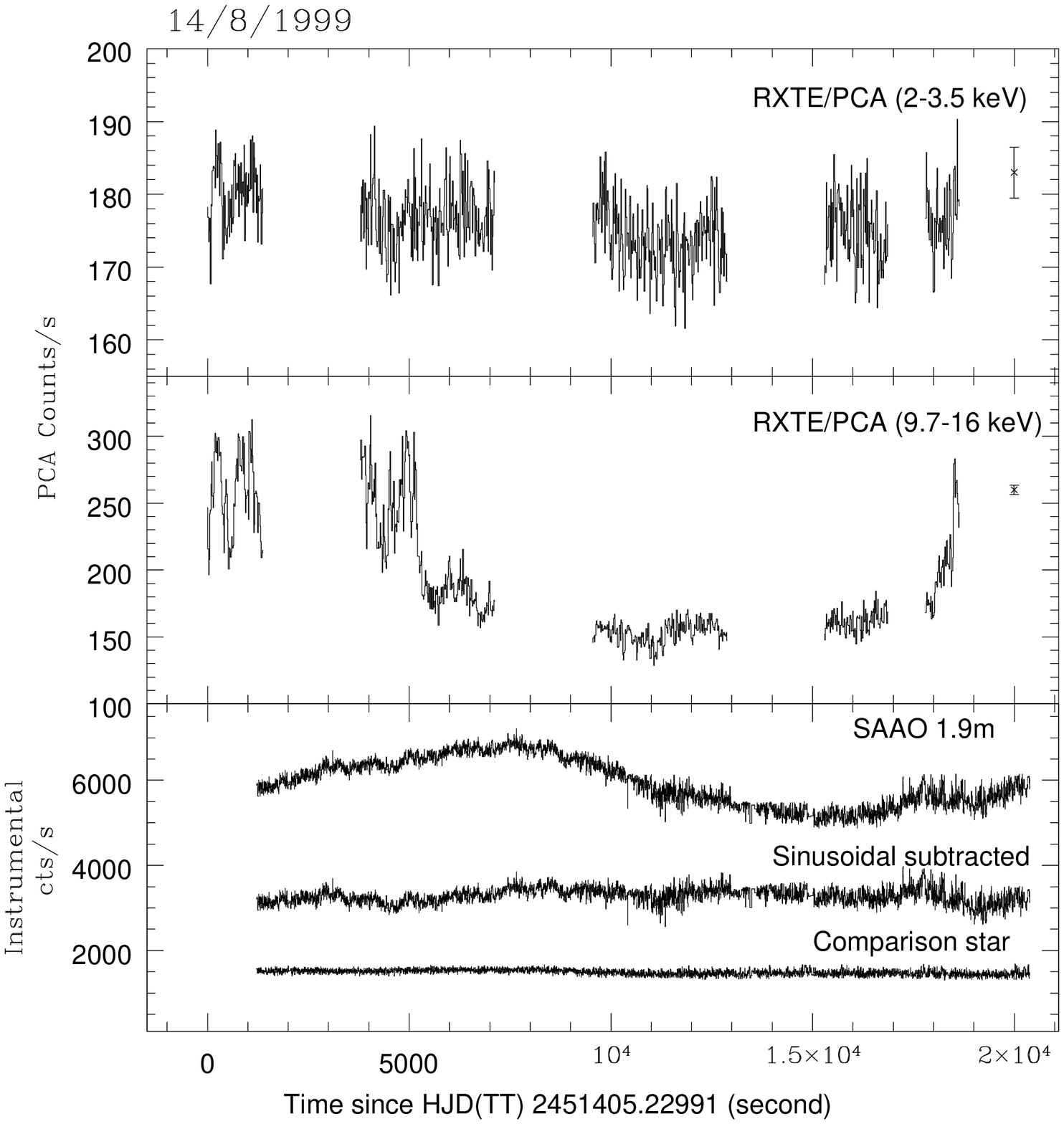,height=14cm}
\caption{Same as Fig.~\ref{fig:lc1}, but for 14/8/99.}
\label{fig:lc4}
\end{center}
\end{figure*}

\begin{figure*}
\begin{center}
\rotatebox{0}{\psfig{file=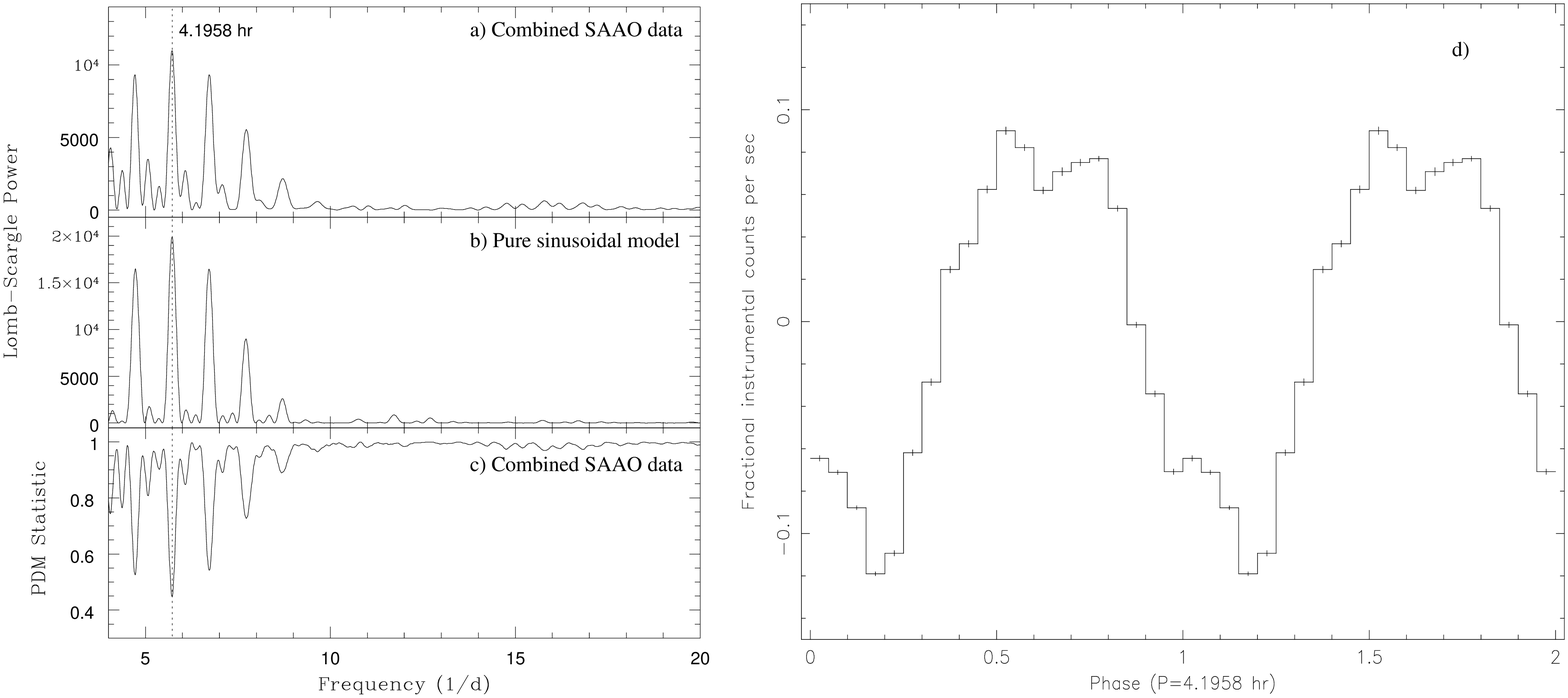,height=8cm,width=15cm}}
\caption{LSP and PDM analysis of the combined SAAO data (left): a)
LSP of GX\,9+9 data; b) Pure sinusoidal at 4.1958~hr and c) PDM of GX\,9+9
data. Right: Folded optical light curve of GX\,9+9 on a period of
4.1958~hr. Two cycles are shown for clarity. Phase zero is arbitrarily
set at the time of the first data point (HJD\,2451402.23164).}
\label{fig:saaolsp}
\end{center}
\end{figure*}

It is very clear from Figs.~\ref{fig:lc1} to \ref{fig:lc4} that substantial
variability is present in the optical light curves of GX\,9+9. In
order to search for any modulation
in the light curves, we used the Lomb-Scargle
periodogram (Lomb 1976; Scargle 1982; hereafter LSP),  
a modification of the discrete Fourier transform which is 
generalised to the case of uneven spacing. We have also employed Phase
Dispersion Minimisation (PDM; \pcite{stellingwerf78}) to check the
results from LSP, as PDM is more sensitive to non-sinusoidal modulation. Before performing period-searching, the
light curves were normalised to fractional count rate in order to eliminate nightly variations. All the light curves were then combined into a single dataset
which was used to calculate the LSP. Fig.~\ref{fig:saaolsp} shows the LSP of
the combined light curve and there is a very strong peak at
$4.1958\pm0.0005$~hr. Also present in the LSP are the 
aliases from this modulation.
This is demonstrated by applying the LSP to a simulated light curve of a pure
sinusoidal modulation at 4.1958~hr and with the same time sampling as the real data (see 
Fig.~\ref{fig:saaolsp}). This shows how the LSP ought to appear if there were only a 4.1958~hr modulation in the data. The principal
features of the LSP are restored when comparing with the real data. A
folded light curve on the 4.1958~hr modulation is shown 
in Fig.~\ref{fig:saaolsp} in which the phase zero is set at the time
of the first data point (HJD\,2451402.23164) and it is highly
sinusoidal with semi-amplitude of 9.8\%. Figure 6 shows the nightly folded
light curves on the 4.1958~hr modulation and the sinusodial modulation is
very stable. It is worth noting that the phase zero roughly corresponds to
minimum light as determined from the data.
A PDM analysis was also
performed which confirmed these results. 
The LSP of several check stars with similar brightness
as the target star were also calculated but none of these show any
modulation at 4.1958~hr, with an upper limit of semi-amplitude
$\sim$~0.17\% . Such small scale 
variations are mainly due to the local atmospheric transparency variations commonly
found at SAAO (e.g. \pcite{homer98}). 
The combined results indicate that the
highly sinusoidal and significant modulation at 4.1958~hr of GX\,9+9 represents the
intrinsic behaviour of the source and is entirely consistent with the
previously reported orbital period (\pcite{schaefer90}) but to a greater degree of accuracy.

\begin{figure*}
\begin{center}
\epsfig{file=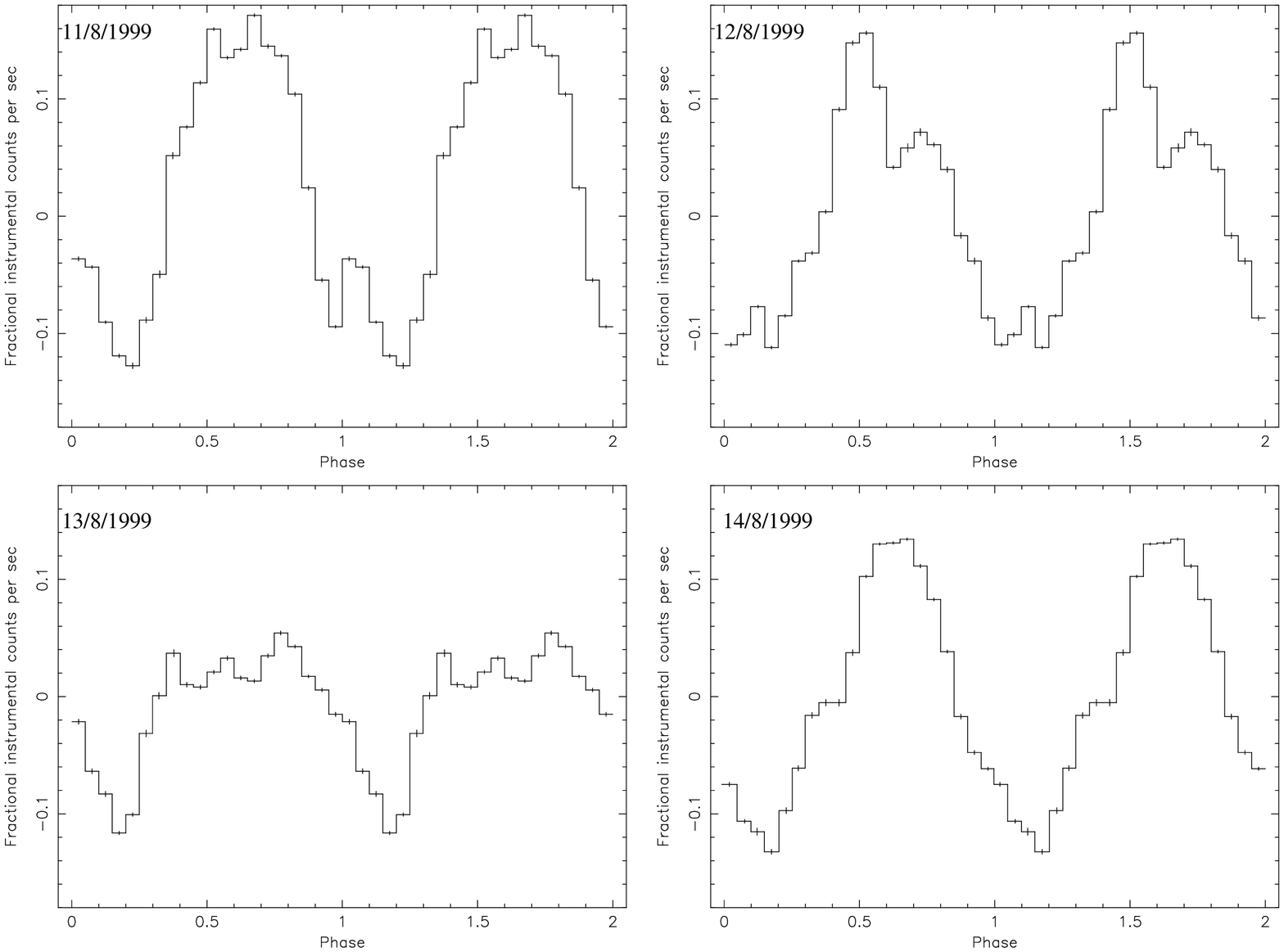,width=15cm}
\caption{Nightly folded optical light curve of GX\,9+9 on a period of
4.1958~hr. Two cycles are shown for clarity. Phase zero is arbitrarily
set at the time of the first data point
(HJD\,2451402.23164). The relatively smaller amplitude of the
August 13, 1999 data is due to the bad seeing near the end of the run
(see also Figure 3).}
\end{center}
\end{figure*}

\begin{figure*}
\begin{center}
\rotatebox{0}{\psfig{file=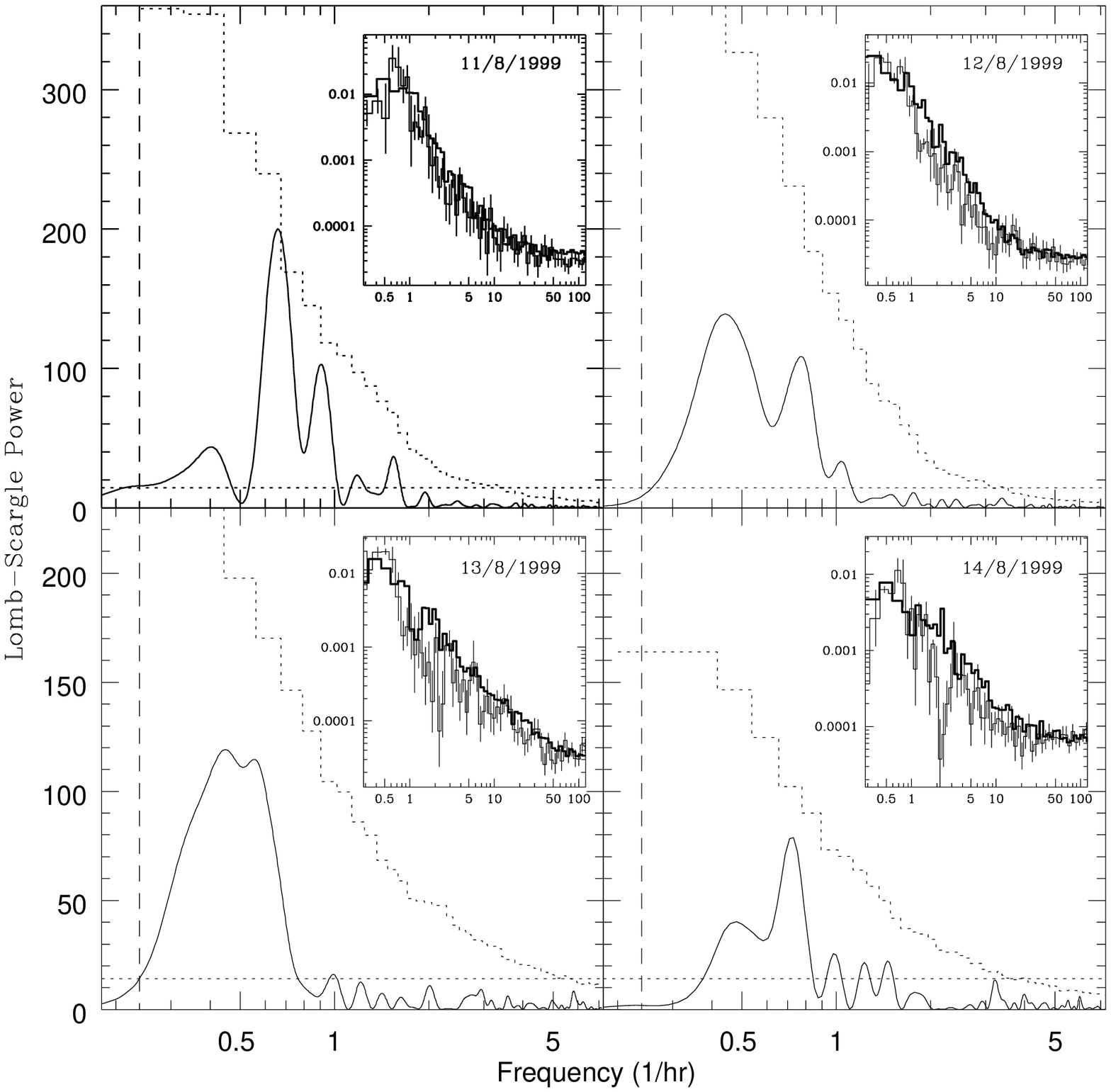,height=14cm}}
\caption{LSP of GX\,9+9 after subtracting the 4.19-hr sinusoidal modulation. The dotted horizontal
line is the 99\% confidence level assuming white noise, while the
dotted histogram is the 99\% confidence level due to red noise. The
4.19-hr orbital period is marked by a vertical dashed line. Insets:
the log-log binned power density spectra (with error bars), with the
average of 200 simulated white-noise plus red-noise models
over-plotted (thick line). Note that the power spectra level off at
high frequency due to the large contribution of white noise.}
\label{fig:4opt}
\end{center}
\end{figure*}

For modulation on shorter timescales ($\sim$~1000~s), we expect any
sub-orbital modulation which is either periodic, or quasi-periodic
will be evident in the LSP or PDM of the data, and should repeat from
night to night. Therefore the LSP and PDM for each
night were calculated in order to search for such variability. Before
calculating both the LSP and PDM, the 4.19-hr orbital modulation was
subtracted (assuming a sinusoid at this period; see Fig.~\ref{fig:lc1}
to \ref{fig:lc4}). In 
Fig.~\ref{fig:4opt}, the
nightly LSP is shown together with the 99\% confidence level in both
white noise (Gaussian noise) and red noise. 
It is apparent that in Fig.~\ref{fig:4opt} those peaks between
0.5--1~hr$^{-1}$ are below the red-noise level. From the log-log binned
power density spectrum (PDS; insets of Fig.~\ref{fig:4opt}), both the
actual and simulated PDS are similar, indicating the variability of
the source on short timescales can be described as the combination of
white and red noise. All the PDS level off at high frequency
($\sim 30$ hr$^{-1}$), suggesting that white noise becomes dominant
between 30 and 900~hr$^{-1}$ (the Nyquist frequency). There is also an indication that an excess appears at $\sim
1.4$~hr from the log-log PDS.
In determining the
white-noise confidence level, we generate Gaussian noise datasets with
the same time intervals and variance as the true data and then perform
the LSP analysis on the resulting datasets. The peak power in each
periodogram (which must be purely due to noise) was then
recorded. This was repeated 10,000 times for good statistics. However,
the noise is not necessarily Gaussian, and can also be
frequency dependent with a higher power at lower frequencies. By using
the above method, strong peaks at the low frequencies might give
misleading results. In order to  take into account such a noise
contribution in the data, we simulate the noise in the form of a power
law, which is also known as red noise
(e.g. \pcite{done92}). Quantitatively, the power spectrum of the
``red-noise'' light curve will be given by $(1/\omega)^{\alpha}$, where
$\omega$ is frequency and $\alpha$ is the slope index. Following an
implementation of the method of \scite{timmer95}, we generate the
simulated light curve with a red-noise power spectrum, together with the
same time sampling and variance as the original data. Finally, we
combine the white-noise simulated light curve as mentioned above, with
the red-noise simulated light curve, by scaling the data points in
each such that the relative contribution of both white and red-noise
components is as specified. Following the method used previously to
determine the white-noise confidence level, we derive the red-noise
confidence level with the combined simulated light curve. Since red
noise is frequency dependent, we need to compute the confidence level
for a set of frequency bins. Hence it looks like a histogram rather
than a horizontal line as for the white-noise level (see Fig.~\ref{fig:4opt}). 
Note that the peak at around 1.4~hr
(0.7~hr$^{-1}$) is just below the red-noise level in the first night
yet it seems to repeat itself in other nights, demonstrating its
quasi-periodic nature.  Hence, the negative result from
the LSP suggests that such variabilities are not strictly
periodic, but are due to some form of quasi-periodic behaviour. 
The 1.4-hr variability may be an intrinsic property of the
source and it may be associated with the mHz QPOs which were also found in
other LMXBs (e.g. Chakrabarty et al. 2001). 

\begin{figure*}
\begin{center}
\rotatebox{0}{\epsfig{file=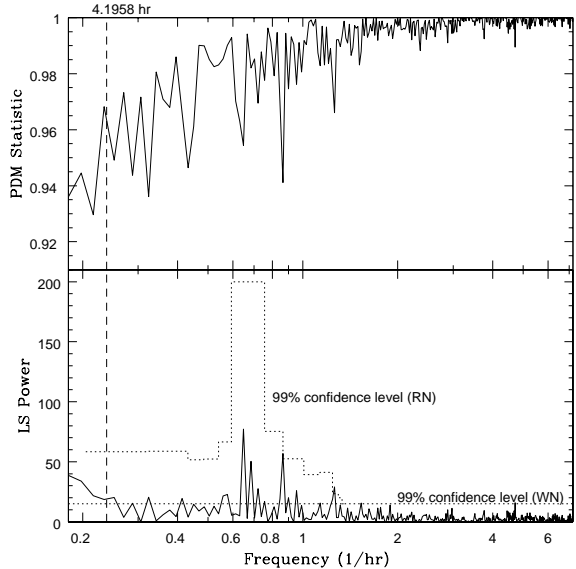,width=8cm}}
\caption{LSP and PDM analysis of the combined \rxte\ data in the 2--3.5~keV band. The
horizontal dotted line is the 99\% confidence level from white noise
(WN), while the dotted histogram is the 99\% confidence level due to
red noise (RN). Note that the sudden jump in the red noise confidence
level is due to the window function of the data.}  
\label{fig:xraylsp}
\end{center}
\end{figure*}

\begin{figure*}
\begin{center}
\rotatebox{0}{\epsfig{file=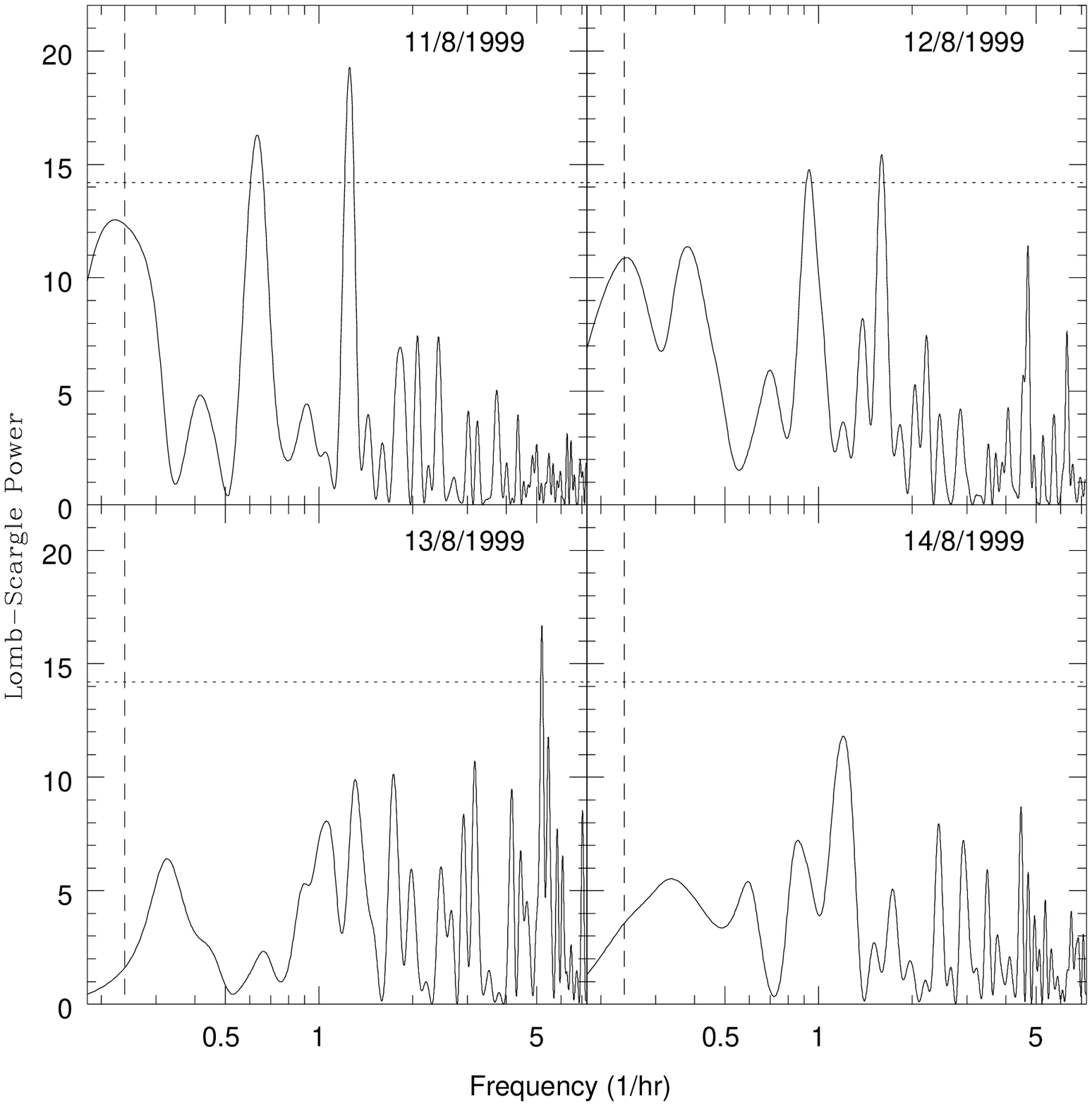,width=8cm}}
\caption{LSP of the nightly \rxte\ data in 2--3.5~keV. The dotted horizontal
line is the 99\% confidence level from white noise, while the vertical
dashed line is the 4.19-hr orbital period.} 
\label{fig:4xraylsp}
\end{center}
\end{figure*}

\subsection{\rxte}
\label{sect:9+9_rxte}
A similar temporal analysis was performed for the \rxte\ data. In order to study the
variability in different energy bands, light curves from two energy
ranges (2--3.5~keV and 9.7--16~keV) were extracted for each
night. From 
Fig.~\ref{fig:lc1}--\ref{fig:lc4}, it is clear that flaring is stronger
at higher energies. In soft X-rays, the count rate roughly remains
at a constant level, while the flares in hard X-rays can be by a
factor of $\sim$~2. The LSP and PDM were calculated from the combined
2--16 keV band pass, as in 
the case of the optical data. The PDM method is particularly important for the X-ray
data since the source changed its spectral state from time to
time (see \S\ref{sect:spec} for details about the spectral behaviour of
the source) and therefore
any modulation would be expected to be non-sinusoidal. Fig.~\ref{fig:xraylsp}
shows the resulting
LSP and PDM for the 2--3.5~keV \rxte\ data. There is no distinct peak
at the optical periodicity of 4.19~hr (marked) and hence the 4.19-hr
X-ray modulation is below the detection threshold in our observations. 
At the 3-$\sigma$ confidence level, upper limits of 
0.9\% and 6.7\% can be set for the semi-amplitude of any 4.19-hr variability in the
2--3.5~keV and 9.7--16~keV bands respectively. The 3.8\% amplitude
seen in {\it HEAO-1} data is below the hard X-ray detection threshold and
it is not implausible that a weaker signal is present.
Note that the
sudden jump at about 
0.6--0.9~hr$^{-1}$ in the 99\% red noise confidence level is due to the window function
arising from the data gaps, typically seen in low-Earth orbit
satellites. Since the red noise is frequency dependent, the confidence
level will be different for different frequency bins and therefore
such analysis also reflects the window function 
(if it exists) in the data.

Variability on shorter timescales is also seen in the LSP but it is
not significant. A period
analysis for 
each night was also performed and the resulting LSPs are shown in
Fig.~\ref{fig:4xraylsp}. There is no significant peak at the 4.19-hr
orbital period which confirms the result from the combined light
curves. On shorter timescales, those marginally 
significant peaks do not repeat more than one night and they are also
comparable to the observing windows of the spacecraft orbit, typically
$\lta 1$~hr due to SAA passages. Hence they are not considered
further.

The X-ray PDS of GX\,9+9 was also calculated for each
night in order to search for kHz QPOs. The PDS was derived from the 125~$\mu$s
time resolution data over 100--4000~Hz. However, the analysis resulted in a
negative finding and an upper limit (95\% confidence) was set of 1.6\%
to the rms amplitude of QPOs with a FWHM of 150~Hz (in 2--60~keV). 

\subsection{Comparison of Light Curves}
\begin{figure*}
\begin{center}
\psfig{file=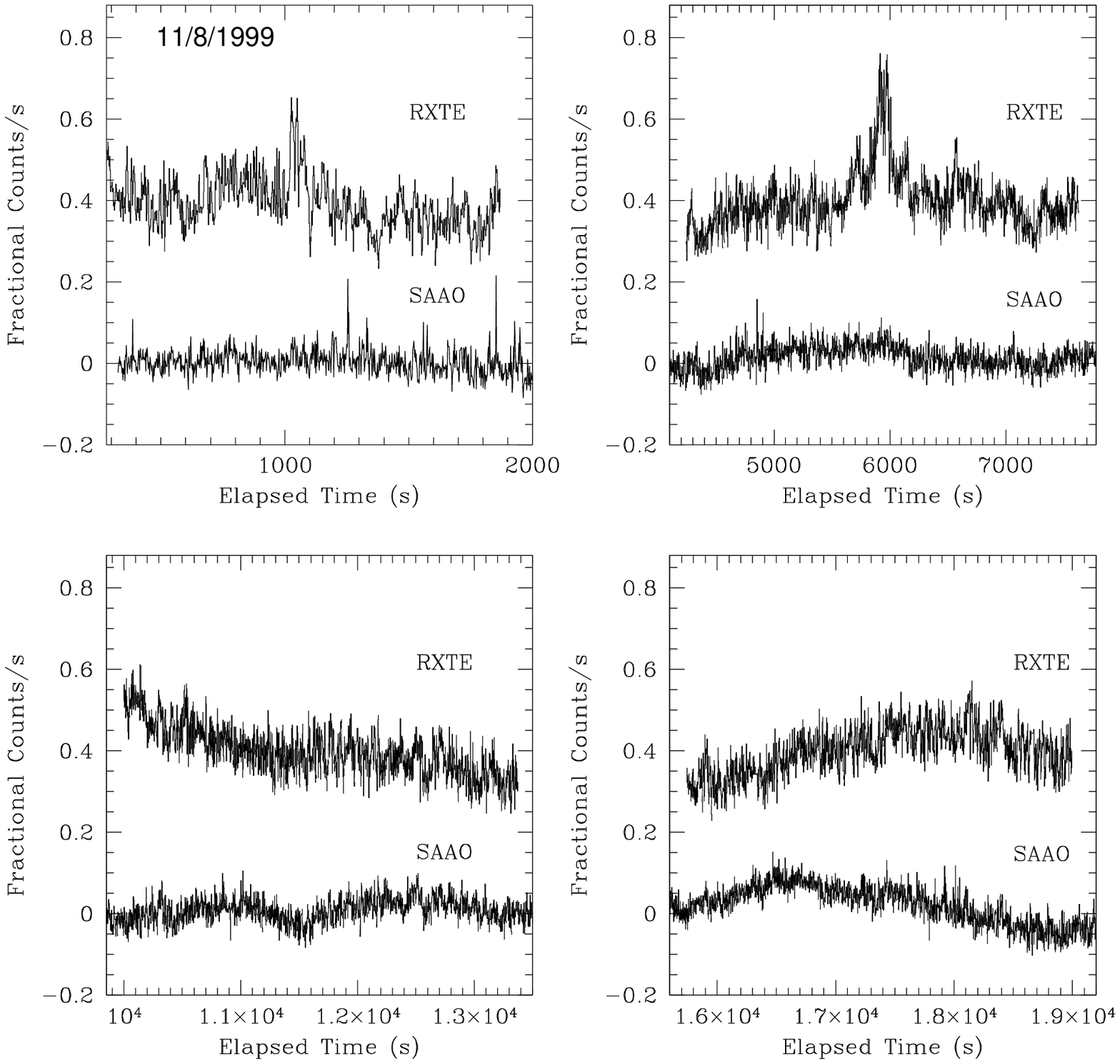,height=15.5cm}
\caption{A close up of the simultaneous X-ray and optical (sinusoidal
  subtracted) light curves of GX\,9+9 on 11/8/99. The timing
  resolution is 1.875~s for X-ray data and 2~s for optical data. The
  count rate for both datasets is normalised to fractional count
  rate.}  
\label{fig:sim_n2}
\end{center}
\end{figure*}

The simultaneous X-ray/optical light curves
(Fig.~\ref{fig:lc1}--\ref{fig:lc4}) exhibit complex variability and it
is interesting to look for any correlation. 
We first applied cross-correlation analysis to compare
the sinusoidal subtracted optical light curves with the
X-ray light curves in different energy bands. We also divided the
light curves into smaller intervals to study the cross-correlation. We repeated this analysis
for all four observations. There is no significant
correlation indicating that the correlated response is very weak. Fig.~\ref{fig:sim_n2} shows a close up of the light
curves
from Fig.~\ref{fig:lc1}, in which there is a prominent X-ray flare in the
second \rxte\ visit. However, there is no immediate response in the
optical. Similar uncorrelated variability can also be found from other
light curves. We have tried to use the transfer function technique as
employed by \scite{kong00a} to investigate the correlations, but the
optical light curves could not be reproduced as a convolution of the
X-ray light curve with a Gaussian transfer function. Hence this suggests
that the optical emission does not significantly respond to X-ray
variability. We note, however, that the presence of uncorrelated events does not
imply the absence of correlated ones.

\section{X-ray Spectroscopy}
\label{sect:spec}

\begin{figure*}
\begin{center}
\rotatebox{-90}{\psfig{file=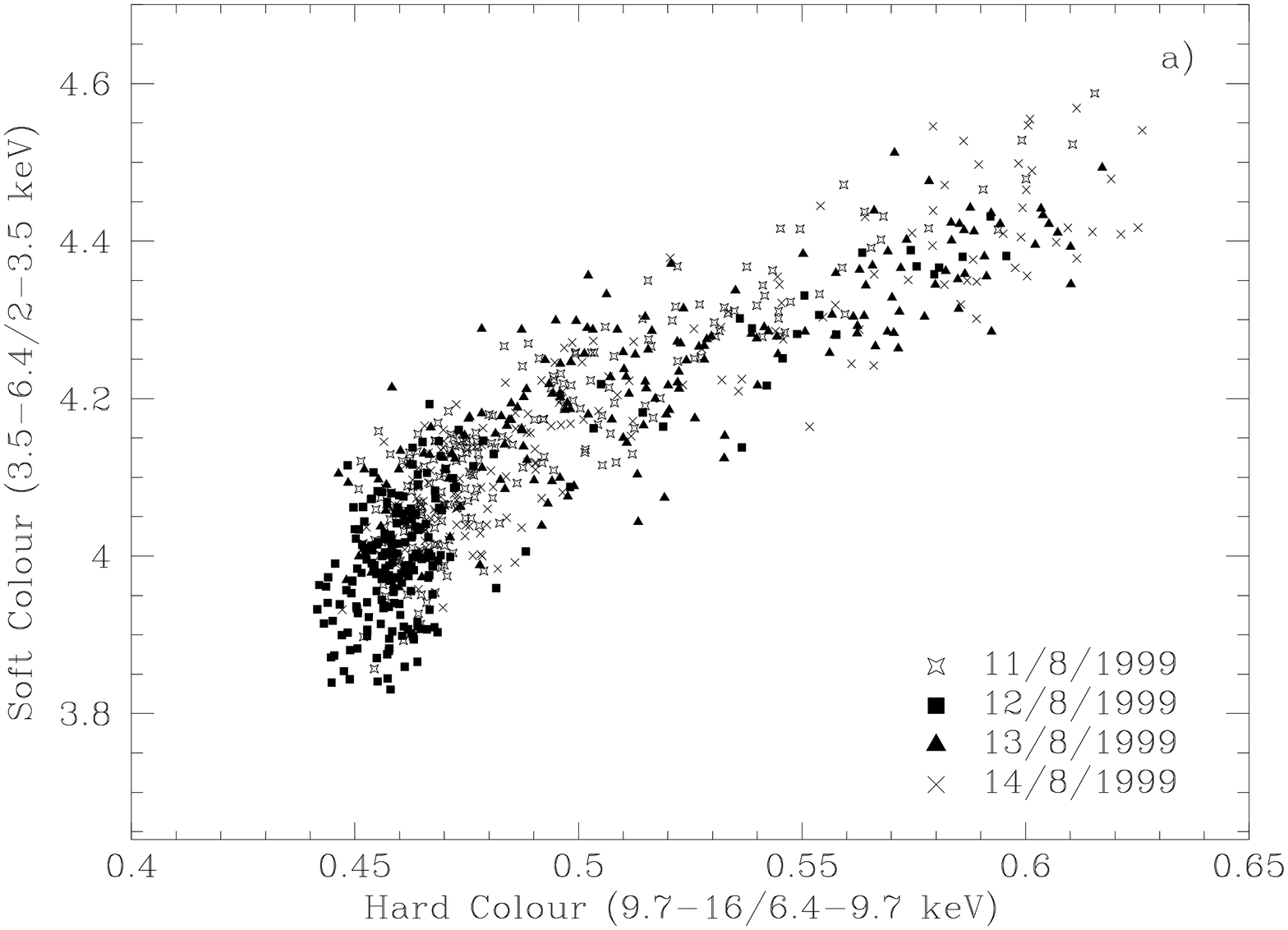,height=8cm,width=7cm}}
\rotatebox{-90}{\psfig{file=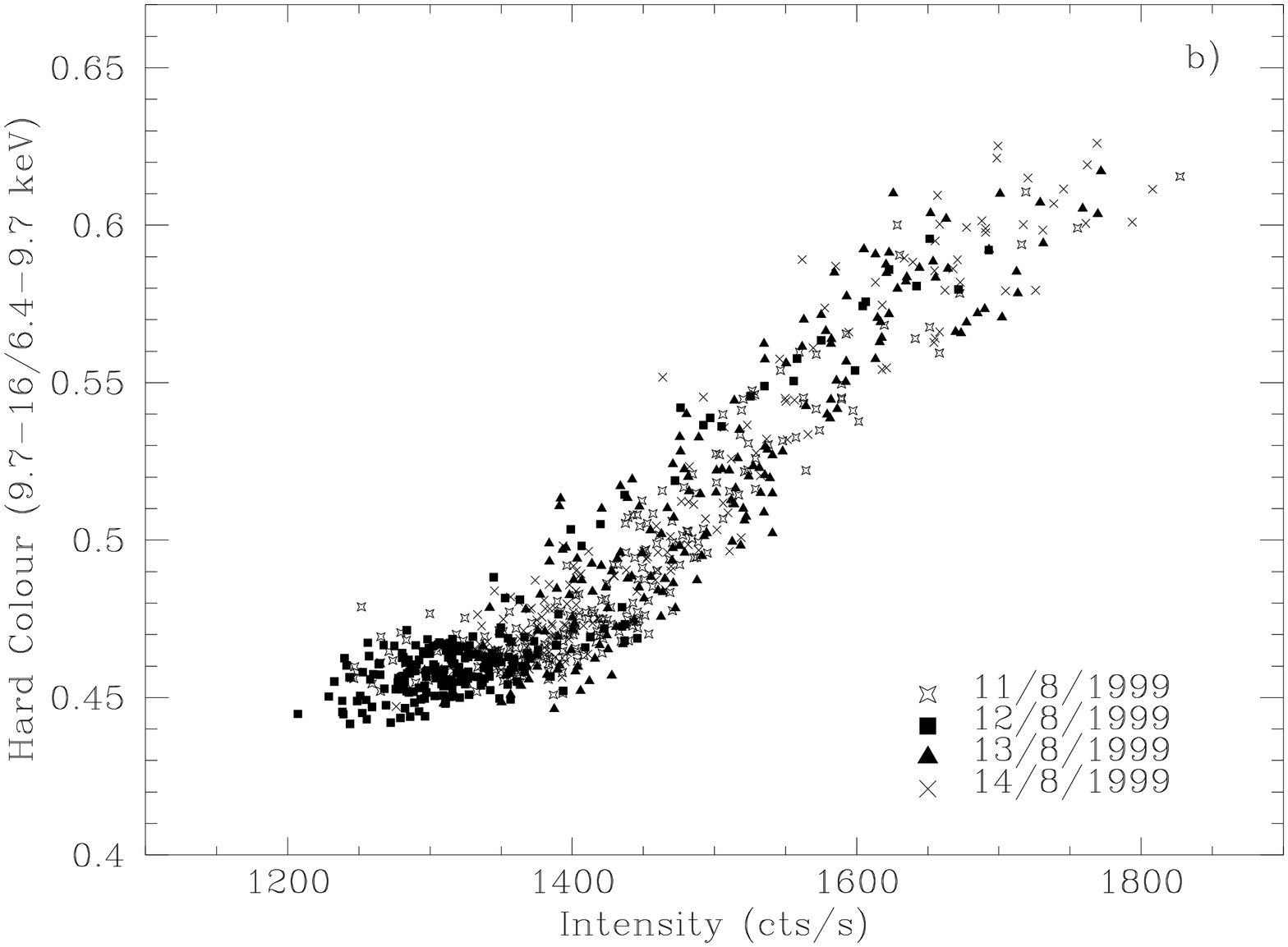,height=8cm,width=7cm}}  
\caption{The colour-colour diagram (a) and hardness-intensity diagram (b) of
all four nights data of GX\,9+9. Each data point is an accumulation of 64~s.}
\label{fig:9+9cd}
\end{center}
\end{figure*}

Colour-colour (CD) and hardness-intensity diagrams (HID) were
derived from the same four energy bands as used by \scite{wijnands98} to
facilitate comparison: 2--3.5, 3.5--6.4, 6.4--9.7 and 9.7--16~keV. The soft and
hard colours are defined as the count rate ratio between 3.5--6.4~keV and
2--3.5~keV and between 9.7--16~keV and 6.4--9.7~keV, respectively. The total
intensity for the HID was taken as the sum of the total count rates (1 PCU) in the
same energy bands as were used in calculating the colours. All the data were
binned to a time resolution of 64 seconds. The CD and HID of all the X-ray
data are shown in Fig.~\ref{fig:9+9cd}. Both the CD and HID of GX\,9+9 suggest
that the source was on the banana branch (Hasinger \& van der Klis 1989) during the \rxte\
observations. Both diagrams show very clear correlation and this is typical for
GX\,9+9 (see e.g. \pcite{schulz89}; Wijnands, van der Klis \& van der Paradijs 1998).

As there is clear indication of correlation between intensity and colour
ratio from the CD and HID, correlated spectral parameters and intensity are
expected, therefore 24 approximately half-hour summed spectra were
derived from the light curves 
(see Table~\ref{tab:cpl}) for further spectral fitting. Based on the
spectral analysis of archival Crab Nebula and pulsar data
(e.g. \pcite{dove98,wilms99,bloser00a,barret00,kong00}), we restrict
our spectral analysis to the energy range of 2.5--20~keV, outside
which the calibration is not reliable. In addition, a 1\% systematic
error was added to account for the uncertainty of the calibration before
the spectral fitting. 

\begin{figure*}
\begin{center}
\rotatebox{90}{\psfig{file=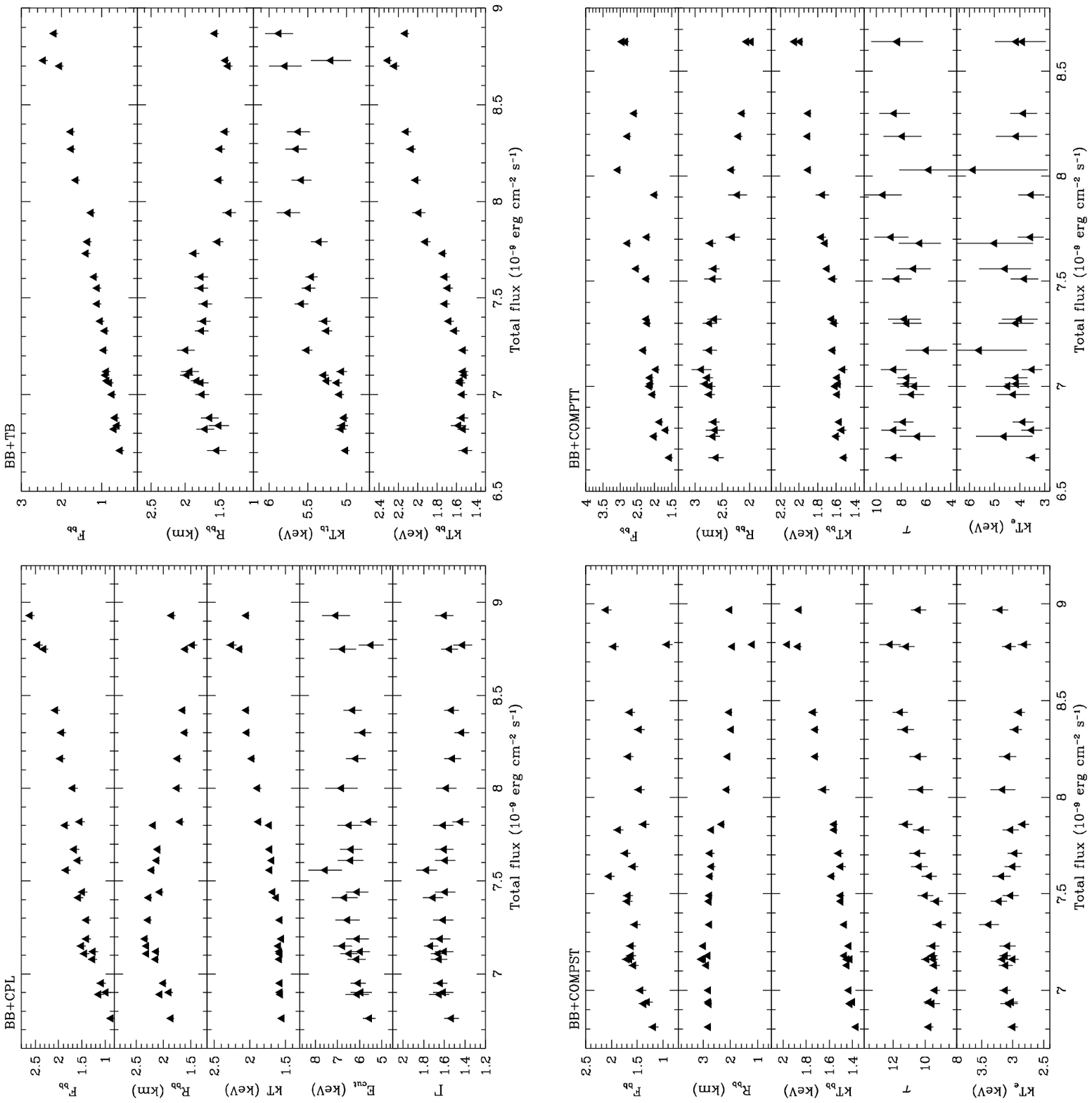,height=17.5cm,width=22.3cm}}
\caption{Variation of spectral parameters for the BB+PL, BB+TB, BB+CompST and BB+CompTT models with the total flux in 2--20~keV.} 
\label{fig:spec}
\end{center}
\end{figure*}

Previous attempts to
fit the X-ray energy spectrum of GX\,9+9 show
that it cannot be fitted well with a single-component model (either power law,
blackbody, cutoff power law or thermal bremsstrahlung; e.g.
\pcite{schulz99}). These single-component models were fitted
first to our \rxte\ data but the results were similarly negative with unacceptable
$\chi_{\nu}^{2}$ ($>2$). Two-component
models were then applied to the spectra. Combinations of blackbody plus
cutoff power law (BB+CPL), blackbody plus thermal bremsstrahlung (BB+TB), and
blackbody plus Comptonization (BB+CompST; \pcite{sunyaev80} and
BB+CompTT; \pcite{titarchuk94}) result in a 
significant improvement to the spectral fits (see 
Tables~\ref{tab:cpl}--\ref{tab:comptt} and
Fig.~\ref{fig:spec}). All models included a column density
$N_H$, which was kept fixed at $2.1\times10^{21}$ cm$^{-2}$ (e.g.
\pcite{church00}) for all the fits since it tended to vary over a wide
range if left as a free parameter. We also note that from an optical
spectroscopic observation \cite{shahbaz96}, the interstellar
absorption features suggest a reddening of $E(B-V)=0.1-0.4$. Hence,
the column density can be derived by using the conversion of
$N_H/10^{22}$cm$^{-2}=0.179 A_V$ \cite{predehl95} and $A_V=3.1 E(B-V)$
\cite{schild77}, suggesting $N_H$ is in the range of $0.6-2.2\times
10^{21}$ cm$^{-2}$. The total flux of GX\,9+9 varies 
between 6.8--8.9 $\times 10^{-9}\fl$, corresponding to
$2-2.7\times10^{37} D_5^2$~\lum\ where $D_5$ is in units of 5~kpc
(e.g. \pcite{church00}) over the four days of our observations. The
best-fit parameters 
of all the models for the 24 spectra are given in
Tables~\ref{tab:cpl}--\ref{tab:comptt}. The effective radius of the blackbody emitting
surface $R_{bb}$ and the ratio of the blackbody flux to the total flux in the
2--20~keV band, $F_{bb}/F_{total}$ are also included. For the BB+CompST and
BB+CompTT models, the Comptonizing $y$-parameter, defined as $y=4kT_e
\tau^2/m_e c^2$ was derived. In Fig.~\ref{fig:spec}, the spectral
parameters of all the models are plotted as a function of the total
(2--20~keV) flux. For all the models,
the blackbody 
temperature is  
about 1.4--2~keV and is always correlated with the total flux. As the total
flux increases, the blackbody component increases from $\sim$~10\% to
40\% while the other component remains roughly
constant. Hence, the increase of the blackbody emission may also be responsible to the flaring activities. 
Such positive correlation also reflects the
correlation between the intensity level and the hardness ratios described in
the HID. The blackbody radius $R_{bb}$ drops slightly from $\sim$~2.5~km to
$\sim$~2~km when the total flux increases. 
 
\begin{table*}
\caption{Spectral fits of GX\,9+9 with the BB+CPL model.}
\label{tab:cpl}
\small
\begin{center}
\begin{tabular}{c c c c c c c c c c c}
\hline
Number & Start (UT) & End (UT) & $\Gamma^a$ & $E_{cut}^b$ & $kT_{bb}^c$ &$F_{tot}^d$ & $F_{bb}/F_{tot}^e$ & $R_{bb}^f$ & $\chi_{\nu}^{2}$ \\
&&&&(keV)&(keV)&&&(km)&(34 dof)\\
\hline
1 & 11 17:22:23& 17:55:59& $1.43\pm0.07$& $5.85\pm0.39$& $2.04\pm0.04$& 8.3 &0.23& $1.61\pm0.06$ & 0.68\\
2 & 11 18:29:03&18:54:19& $1.59\pm0.10$&$6.41\pm0.58$&$1.69\pm0.02$&7.61&0.21& $2.12\pm0.05$&0.88\\
3& 11 19:02:23&19:35:59&$1.44\pm0.08$&$5.60\pm0.38$&$1.88\pm0.04$&7.82&0.20&$1.69\pm0.07$&0.67\\
4&11 20:25:43&21:15:59&$1.65\pm0.08$&$6.13\pm0.41$&$1.59\pm0.02$&7.08&0.18&$2.14\pm0.05$&1.08\\
5&11 21:15:59&22:39:19&$1.66\pm0.07$&$6.48\pm0.39$&$1.58\pm0.01$&7.11&0.21&$2.31\pm0.03$& 0.60\\
6&11 23:12:23&23:45:59&$1.71\pm0.10$&$6.68\pm0.59$&$1.63\pm0.02$&7.41&0.21&$2.27\pm0.06$&0.91\\
7&12 14:28:15&18:01:51&$1.65\pm0.10$&$6.10\pm0.53$&$1.57\pm0.02$&6.89&0.16&$2.07\pm0.05$&0.77\\
8&12 18:34:55&19:25:11&$1.64\pm0.07$&$6.06\pm0.35$&$1.58\pm0.02$&6.95&0.16&$2.00\pm0.05$&0.68\\
9&12 19:58:15&20:31:51&$1.61\pm0.10$&$5.93\pm0.49$&$1.58\pm0.03$&6.90&0.14&$1.90\pm0.07$&0.88\\
10&12 20:31:52&21:05:11&$1.60\pm0.09$&$5.97\pm0.43$&$1.59\pm0.02$&7.12&0.18&$2.13\pm0.05$&0.91\\
11&12 21:38:15&22:45:11&$1.53\pm0.07$&$5.56\pm0.29$&$1.55\pm0.02$&6.76&0.13&$1.86\pm0.05$&0.79\\
12&13 17:26:23&17:59:59&$1.64\pm0.10$&$6.09\pm0.52$&$1.56\pm0.02$&7.19&0.19&$2.34\pm0.06$&1.01\\
13&13 18:16:23&18:49:59&$1.42\pm0.09$&$5.48\pm0.57$&$2.26\pm0.07$&8.77&0.28&$1.47\pm0.09$&0.87\\
14&13 18:50:00&19:39:59&$1.53\pm0.07$&$6.31\pm0.41$&$2.05\pm0.04$&8.42&0.25&$1.65\pm0.06$&0.5\\
15&13 19:56:23&20:29:59&$1.59\pm0.10$&$6.11\pm0.51$&$1.68\pm0.03$&7.44&0.20&$2.07\pm0.07$&0.86\\
16&13 20:29:59&20:54:59&$1.61\pm0.10$&$6.46\pm0.55$&$1.73\pm0.02$&7.80&0.24&$2.19\pm0.05$&1.21\\
17&13 21:36:23&22:09:59&$1.52\pm0.08$&$6.16\pm0.45$&$1.97\pm0.04$&8.16&0.24&$1.74\pm0.07$&0.86\\
18&13 22:09:59&23:00:00&$1.77\pm0.10$&$7.56\pm0.75$&$1.72\pm0.02$&7.56&0.24&$2.21\pm0.05$&1.07\\
19&14 17:25:35&17:59:11&$1.60\pm0.09$&$7.07\pm0.63$&$2.05\pm0.04$&8.93&0.29&$1.85\pm0.07$&0.82\\
20&14 18:23:55&18:50:51&$1.55\pm0.08$&$6.76\pm0.60$&$2.14\pm0.04$&8.75&0.27&$1.60\pm0.06$&0.78\\
21&14 18:57:15&19:39:11&$1.60\pm0.09$&$6.39\pm0.51$&$1.72\pm0.02$&7.67&0.22&$2.10\pm0.05$&0.64\\
22&14 19:55:35&21:02:31&$1.73\pm0.07$&$6.76\pm0.42$&$1.60\pm0.01$&7.15&0.21&$2.31\pm0.03$&0.7\\
23&14 21:35:35&22:09:11&$1.61\pm0.10$&$6.55\pm0.56$&$1.58\pm0.02$&7.29&0.19&$2.28\pm0.06$&0.9\\
24&14 22:17:15&22:42:31&$1.58\pm0.10$&$6.82\pm0.74$&$1.89\pm0.05$&8.00&0.21&$1.75\pm0.09$&0.93\\ 
\hline
\\
\multicolumn{10}{l}{$^a$ Photon index}\\
\multicolumn{10}{l}{$^b$ Cutoff energy}\\
\multicolumn{10}{l}{$^c$ Blackbody temperature}\\
\multicolumn{10}{l}{$^d$ Total flux in 2--20~keV ($10^{-9}$\fl)}\\
\multicolumn{10}{l}{$^e$ Ratio of blackbody flux to total flux, 2--20~keV}\\
\multicolumn{10}{l}{$^f$ Blackbody radius, assuming a distance of 5~kpc}\\
\end{tabular}
\end{center}
\normalsize
\end{table*}

\begin{table*}
\caption{Spectral fits of GX\,9+9 with the BB+TB model.}
\label{tab:tb}
\begin{center}
\begin{tabular}{c c c c c c c} 
\hline
Number & $kT_{bb}$ &$kT_{tb}^a$&$F_{tot}$ & $F_{bb}/F_{tot}$ & $R_{bb}$ & $\chi_{\nu}^{2}$ \\
&(keV)&(keV)&&&(km)&(35 dof)\\
\hline
1&$2.07\pm0.05$&$5.65\pm0.14$&8.27&0.21&$1.49\pm0.07$&0.69\\
2&$1.69\pm0.05$&$5.49\pm0.09$&7.55&0.15&$1.77\pm0.10$&0.92\\
3&$1.92\pm0.05$&$5.35\pm0.11$&7.79&0.17&$1.52\pm0.08$&0.66\\
4&$1.54\pm0.05$&$5.09\pm0.06$&7.00&0.11&$1.75\pm0.10$&1.25\\ 
5&$1.52\pm0.04$&$5.29\pm0.06$&7.10&0.13&$1.98\pm0.10$&0.72\\
6&$1.62\pm0.05$&$5.25\pm0.07$&7.33&0.13&$1.76\pm0.10$&1.06\\
7&$1.53\pm0.06$&$5.07\pm0.07$&6.82&0.10&$1.71\pm0.13$&0.90\\
8&$1.54\pm0.06$&$5.03\pm0.05$&6.88&0.10&$1.64\pm0.13$&0.80\\
9&$1.58\pm0.08$&$5.05\pm0.07$&6.84&0.09&$1.51\pm0.15$&0.92\\ 
10&$1.56\pm0.05$&$5.12\pm0.06$&7.06&0.11&$1.77\pm0.11$&1.06\\
11&$1.51\pm0.07$&$5.01\pm0.05$&6.71&0.08&$1.54\pm0.14$&0.84\\
12&$1.53\pm0.05$&$5.06\pm0.07$&7.12&0.12&$1.93\pm0.13$&1.11\\
13&$2.31\pm0.04$&$5.20\pm0.26$&8.73&0.28&$1.41\pm0.05$&0.83\\
14&$2.12\pm0.05$&$5.62\pm0.15$&8.36&0.21&$1.42\pm0.07$&0.58\\
15&$1.68\pm0.05$&$5.28\pm0.08$&7.38&0.14&$1.73\pm0.10$&0.95\\
16&$1.74\pm0.04$&$4.48\pm0.09$&7.73&0.18&$1.87\pm0.08$&1.30\\
17&$2.02\pm0.05$&$5.58\pm0.13$&8.11&0.20&$1.51\pm0.07$&0.91\\
18&$1.72\pm0.05$&$5.58\pm0.09$&7.47&0.15&$1.71\pm0.10$&1.42\\
19&$2.13\pm0.04$&$5.87\pm0.18$&8.87&0.25&$1.57\pm0.06$&0.92\\
20&$2.24\pm0.05$&$5.79\pm0.21$&8.70&0.24&$1.37\pm0.06$&0.86\\
21&$1.72\pm0.05$&$5.45\pm0.08$&7.61&0.16&$1.77\pm0.10$&0.75\\
22&$1.56\pm0.04$&$5.25\pm0.06$&7.07&0.12&$1.83\pm0.09$&0.92\\
23&$1.53\pm0.05$&$5.51\pm0.07$&7.23&0.13&$1.99\pm0.13$&1.03\\
24&$1.99\pm0.07$&$5.75\pm0.15$&7.94&0.16&$1.36\pm0.10$&0.95\\
\hline
\\
\multicolumn{7}{l}{$^a$ Thermal bremsstrahlung temperature}\\
\end{tabular}
\end{center}
\normalsize
\end{table*}

\begin{table*}
\caption{Spectral fits of GX\,9+9 with the BB+CompST model.}
\label{tab:compst}
\begin{center}
\begin{tabular}{c c c c c c c c c} 
\hline
Number & $kT_{bb}$ & $kT_{e}^a$ & $\tau^b$& $y^c$& $F_{tot}$ & $F_{bb}/F_{tot}$ & $R_{bb}$ & $\chi_{\nu}^{2}$ \\
&(keV)&(keV)&&&&&(km)&(34 dof)\\
\hline
1&$1.72\pm0.03$&$2.95\pm0.10$&$11.22\pm0.52$&2.90&8.35&0.17&$1.97\pm0.07$&0.69\\
2&$1.50\pm0.04$&$3.00\pm0.13$&$10.36\pm0.53$&2.50&7.64&0.21&$2.69\pm0.14$&0.90\\
3&$1.56\pm0.03$&$2.83\pm0.09$&$11.22\pm0.42$&2.78&7.86&0.17&$2.31\pm0.09$&0.69\\
4&$1.45\pm0.02$&$3.11\pm0.11$&$9.40\pm0.34$&2.15&7.13&0.22&$2.87\pm0.08$&1.17\\
5&$1.45\pm0.02$&$3.16\pm0.09$&$9.48\pm0.28$&2.22&7.16&0.24&$2.98\pm0.08$&0.62\\
6&$1.50\pm0.02$&$3.22\pm0.13$&$9.27\pm0.39$&2.16&7.46&0.23&$2.78\pm0.07$&0.96\\
7&$1.42\pm0.03$&$3.06\pm0.14$&$9.55\pm0.48$&2.18&6.93&0.20&$2.79\pm0.12$&0.88\\
8&$1.43\pm0.02$&$3.12\pm0.09$&$9.37\pm0.29$&2.14&7.00&0.20&$2.81\pm0.08$&0.75\\
9&$1.40\pm0.02$&$3.02\pm0.10$&$9.75\pm0.34$&2.24&6.94&0.19&$2.80\pm0.08$&0.94\\
10&$1.42\pm0.02$&$3.00\pm0.10$&$9.87\pm0.37$&2.28&7.16&0.23&$3.06\pm0.09$&1.13\\
11&$1.37\pm0.02$&$3.00\pm0.08$&$9.77\pm0.29$&2.24&6.81&0.17&$2.80\pm0.08$&0.91\\
12&$1.43\pm0.02$&$3.08\pm0.13$&$9.51\pm0.42$&2.18&7.23&0.22&$2.99\pm0.09$&1.08\\
13&$1.96\pm0.02$&$2.81\pm0.11$&$12.18\pm0.68$&3.60&8.79&0.10&$1.20\pm0.02$&0.87\\
14&$1.74\pm0.04$&$2.89\pm0.09$&$11.54\pm0.48$&3.00&8.44&0.19&$2.04\pm0.10$&0.58\\
15&$1.50\pm0.03$&$3.03\pm0.13$&$9.97\pm0.48$&2.35&7.49&0.22&$2.77\pm0.11$&0.96\\
16&$1.56\pm0.03$&$3.03\pm0.13$&$10.23\pm0.52$&2.47&7.83&0.24&$2.70\pm 0.10$&1.32\\
17&$1.72\pm0.03$&$3.08\pm0.14$&$10.43\pm0.55$&2.62&8.21&0.20&$2.10\pm0.07$&0.91\\
18&$1.58\pm0.03$&$3.18\pm0.15$&$9.73\pm0.48$&2.35&7.59&0.27&$2.75\pm0.10$&1.35\\
19&$1.86\pm0.02$&$3.20\pm0.13$&$10.41\pm0.48$&2.71&8.97&0.23&$2.01\pm0.04$&0.86\\
20&$1.87\pm0.03$&$3.06\pm0.11$&$11.15\pm0.49$&2.97&8.78&0.22&$1.93\pm0.06$&0.83\\
21&$1.52\pm0.04$&$2.96\pm0.12$&$10.46\pm0.50$&2.53&7.71&0.22&$2.74\pm0.15$&0.80\\
22&$1.47\pm0.02$&$3.12\pm0.09$&$9.53\pm0.30$&2.22&7.18&0.23&$2.83\pm0.07$&0.82\\
23&$1.47\pm0.02$&$3.38\pm0.16$&$9.12\pm0.43$&2.20&7.34&0.21&$2.76\pm0.07$&1.00\\
24&$1.65\pm0.05$&$3.16\pm0.20$&$10.27\pm0.77$&2.61&8.04&0.18&$2.14\pm0.13$&0.93\\
\hline
\\
\multicolumn{9}{l}{$^a$ Electron temperature of the Comptonizing cloud}\\
\multicolumn{9}{l}{$^b$ Optical depth of the Comptonizing cloud}\\
\multicolumn{9}{l}{$^c$ Comptonization parameter $y=4kT_e \tau^2/m_e c^2$}\\
\end{tabular}
\end{center}
\normalsize
\end{table*}

\begin{table*}
\caption{Spectral fits of GX\,9+9 with the BB+CompTT model.}
\label{tab:comptt}
\small
\begin{center}
\begin{tabular}{c c c c c c c c c c} 
\hline
Number & $kT_{bb}$ & $T_0^a$& $kT_{e}$ & $\tau$& $y$& $F_{tot}$ & $F_{bb}/F_{tot}$ & $R_{bb}$ & $\chi_{\nu}^{2}$ \\
&(keV)&(keV)&(keV)&&&&&(km)&(33 dof)\\
\hline
1&$1.91\pm0.03$&$0.55\pm0.02$&$4.12\pm0.82$&$7.90\pm1.54$&2.00&8.19&0.34& $2.20\pm0.07$&0.62\\
2&$1.64\pm0.05$&$0.50\pm0.03$&$3.81\pm0.55$&$8.38\pm1.20$&2.04&7.51&0.30& $2.67\pm0.16$&0.86\\
3&$1.76\pm0.05$&$0.53\pm0.02$&$3.56\pm0.51$&$8.81\pm1.37$&2.16&7.71&0.29& $2.31\pm0.13$&0.65\\
4&$1.59\pm0.03$&$0.51\pm0.02$&$4.26\pm0.66$&$7.16\pm1.00$&1.71&6.96&0.30& $2.73\pm0.10$&1.00\\
5&$1.58\pm0.03$&$0.50\pm0.02$&$4.13\pm0.45$&$7.58\pm0.78$&1.85&7.01&0.30& $2.81\pm0.10$&0.56\\
6&$1.62\pm0.04$&$0.50\pm0.02$&$4.14\pm0.70$&$7.53\pm1.18$&1.83&7.30&0.30& $2.73\pm0.13$&0.86\\
7&$1.60\pm0.04$&$0.53\pm0.02$&$4.61\pm1.13$&$6.67\pm1.44$&1.60&6.76&0.30& $2.67\pm0.13$&0.64\\
8&$1.57\pm0.03$&$0.51\pm0.02$&$3.86\pm0.41$&$7.81\pm0.80$&1.84&6.83&0.27& $2.65\pm0.10$&0.66\\
9&$1.54\pm0.05$&$0.50\pm0.02$&$3.52\pm0.41$&$8.59\pm1.00$&2.03&6.79&0.25& $2.63\pm0.17$&0.89\\
10&$1.61\pm0.03$&$0.52\pm0.02$&$4.48\pm0.87$&$6.90\pm1.20$&1.67&7.00&0.31& $2.72\pm0.10$&0.79\\
11&$1.52\pm0.04$&$0.51\pm0.02$&$3.48\pm0.26$&$8.62\pm0.68$&2.02&6.66&0.24& $2.61\pm0.14$&0.82\\
12&$1.53\pm0.05$&$0.49\pm0.02$&$3.51\pm0.40$&$8.60\pm1.03$&2.03&7.08&0.28& $2.88\pm0.18$&0.99\\
13&$2.05\pm0.03$&$0.55\pm0.03$&$3.88\pm0.92$&$8.32\pm2.09$&2.10&8.64&0.34& $1.97\pm0.05$&0.77\\
14&$1.90\pm0.03$&$0.52\pm0.02$&$3.84\pm0.54$&$8.55\pm1.24$&2.19&8.30&0.31& $2.14\pm0.07$&0.50\\
15&$1.65\pm0.04$&$0.52\pm0.02$&$4.00\pm0.71$&$7.74\pm1.32$&1.87&7.32&0.31& $2.64\pm0.13$&0.78\\
16&$1.72\pm0.03$&$0.53\pm0.02$&$5.00\pm1.53$&$6.50\pm1.69$&1.65&7.68&0.36& $2.71\pm0.10$&1.14\\
17&$1.90\pm0.03$&$0.56\pm0.02$&$5.86\pm2.98$&$5.76\pm2.42$&1.52&8.03&3.06& $2.33\pm0.07$&0.78\\
18&$1.75\pm0.03$&$0.53\pm0.02$&$10.41\pm14.95$&$3.70\pm4.47$&1.11&7.42&0.37&$2.61\pm0.09$&0.78\\
19&$1.96\pm0.03$&$0.53\pm0.02$&$4.74\pm1.36$&$7.24\pm1.85$&1.94&8.81&0.38& $2.30\pm0.07$&0.79\\
20&$1.99\pm0.03$&$0.53\pm0.02$&$4.12\pm0.86$&$8.26\pm1.71$&2.20&8.64&0.33& $2.05\pm0.06$&0.75\\
21&$1.70\pm0.03$&$0.52\pm0.02$&$4.58\pm1.03$&$7.02\pm1.39$&1.76&7.56&0.33& $2.65\pm0.10$&0.58\\
22&$1.59\pm0.03$&$0.49\pm0.02$&$4.14\pm0.46$&$7.53\pm0.76$&1.83&7.04&0.30& $2.77\pm0.10$&0.70\\
23&$1.64\pm0.04$&$0.54\pm0.02$&$5.62\pm1.90$&$5.98\pm1.66$&1.57&7.17&0.32& $2.73\pm0.13$&0.79\\
24&$1.75\pm0.07$&$0.49\pm0.04$&$3.54\pm0.51$&$9.45\pm1.50$&2.47&7.91&0.25& $2.22\pm0.17$&0.92\\ 
\hline
\\
\multicolumn{10}{l}{$^a$ Temperature of seed photons}\\
\end{tabular}
\end{center}
\normalsize
\end{table*}

For the BB+CPL model, the photon index is roughly at $\sim$~1.6 with no
indication of any correlation with the total flux. Meanwhile, the cutoff energy
$E_{cut}$ varies from 5.5~keV to 7~keV. Such results indicate that the
contribution of the CPL component remains roughly constant as the total flux
increases, while the BB component becomes the major driving force for the
spectral evolution of the source. For the BB+TB model, the thermal bremsstrahlung
temperature varies from 5~keV to 6~keV and is correlated with the total flux.

For the BB+CompST, the optical depth $\tau$ is moderate at about 9--12 and might
be roughly anti-correlated with the blackbody contribution. The electron
temperature is very steady at $\sim$~3~keV although small variations can be seen and are also anti-correlated with the optical depth. The BB+CompTT results are very similar to BB+CompST but with larger error bars in $\tau$ and $kT_e$. The additional parameter in CompTT, the seed photon temperature ($T_0$), is steady at $\sim$~0.5~keV. Also $\tau$ in CompTT is
slightly smaller than that of the CompST while $kT_e$ is larger than that of the CompST. 

We also investigate the effect of replacing the BB component with a multi-colour disc blackbody model
(DBB; \pcite{mitsuda84}) in all models. All the fits were
acceptable but resulted in a large colour temperature for the inner disc ($kT_{in}\gta 3$~keV) and an unfeasibly small inner disc radius ($R_{in}=1$~km). The
error bars for the other component (i.e. CPL, CompST and CompTT) were also
large compared to the BB models. Therefore the DBB model seems to be
unrealistic.

\section{Archival X-ray Observations}
\begin{figure}
\begin{center}
\rotatebox{0}{\psfig{file=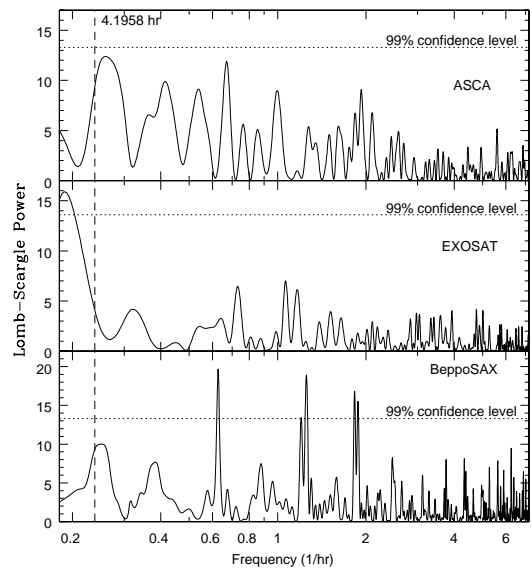,height=8cm}}
\caption{LSP of the \asca, \exosat\ and \sax\ observations of GX\,9+9
  (see Table~\ref{tab:archlog}). The dotted horizontal line is the
  99\% confidence level, while the vertical dashed line is the 4.19-hr
  orbital period. No significant peak at the orbital period is seen in
  either plot. Note that the peaks in the \sax\ LSP are due to the
  window function arising from the data gaps, typically seen in
  low-Earth orbit satellites.} 
\label{fig:asca}
\end{center}
\end{figure}

If the 4.19-hr
modulation is orbital in nature, it should be very stable (as in the
optical). In order to study
the X-ray modulation on this timescale, light curves from the public data
archives were obtained and only observations longer than 15~ks
($\gta$~4.2~hr) were considered. Table~\ref{tab:archlog} shows all the
relevant observations with the date and duration of the
observations. An $\sim$~17.4-hr nearly uninterrupted \exosat\ observation and
an $\sim$~13-hr \asca\ observation were extracted from the archival 
database. In addition, we also analysed our \sax\ data taken in early 
2000. Both the LSP
and PDM were applied in the same way to search for any periodicity.
Fig.~\ref{fig:asca} shows the results from the LSP for the \asca\ 
GIS, {\it EXOSAT} ME  and \sax\ LECS data. Clearly no period at 4.19~hr is detected. In
addition, variability on shorter timescales is not seen in the LSP, which
confirms our \rxte\ results described above.

\begin{table*}
\caption{Archival X-ray observations of GX\,9+9.}
\label{tab:archlog}
\begin{center}
\begin{tabular}{l c c c c c} \hline
Date&Satellite& Energy & Time span & On-source time & Remark\\
    &         &  Range (keV)& ks / hr  &   ks / hr & \\ 
\hline
1983 Sept 27& {\it EXOSAT} & 0.05--2 & 62.6 / 17.4& 51.7 / 14.4 & Nearly uninterrupted\\
1994 March 25 & \asca & 0.5--10 & 47.2 / 13 & 23.7 / 6.6& \\
2000 April 8 & \sax & 0.1--10 &100 / 27.8 & 23.6 / 6.4&\\
\hline
\end{tabular}
\end{center}
\end{table*}

\section{Discussion}
\label{sect:9+9discussion}

\subsection{Orbital Modulation}
We have observed the luminous LMXB GX\,9+9 simultaneously with X-ray and optical telescopes. In the optical, the source was seen at $B=16.96\pm0.05$ and $V=16.79\pm0.05$, which is very consistent with previous observations \cite{schaefer90}. The high quality optical light curves of GX\,9+9 presented here confirm the
4.19-hr periodicity found previously \cite{schaefer90}. The stability of
this period also shows that it is the orbital period of the system. Unlike in
the optical,
no significant X-ray modulation on the 4.19~hr period was
detected in the four-night \rxte\ observations and other archival
X-ray observations (see \S5). This is somewhat
inconsistent with the previous {\it HEAO-1} results \cite{hertz88}, in which
a 3.8\% modulation was found. In fact, the {\it HEAO-1} data showed
large scatter (4.6\% residual error) and the light curve was based on
74 data points, typically $\sim$~1.9~hr 
apart and spread over 5.91 days. Their detected 4.19~hr X-ray
modulation is only marginally significant at the 99\% confidence
level. Comparing the two observations from {\it HEAO-1}, the 4.19-hr
X-ray modulation is only found in one of them. Since the source
often has flaring and short-term variation ($<$~1~hr) especially in
hard X-rays 
($>$~10~keV; see Fig.~\ref{fig:lc1}--\ref{fig:lc4}), such
activities might cause a false alarm in determining
the period and account for the large errors if the time resolution is
poor. Hence, these results should be treated with caution. The \rxte\
data and other archival observations reported here 
provide much higher time resolution and energy-selected light curves to
study the 4.19-hr modulation of this source in greater detail. Both LSP and
PDM analysis show no indication of variability on this timescale.
However, it is possible that an X-ray periodicity might be detectable 
from time to time as a result of changes in the accretion geometry (such 
as the height of the disc rim), and it should be noted that the X-ray 
period was discovered prior to the optical modulation.  Consequently, the 
agreement between the two periodicities means that the original X-ray 
modulation should not be discarded.

The lack of X-ray modulation and eclipses or absorption dips on the
orbital period is common amongst 
LMXBs and indicates that the inclination angle of the system must be low
($< 70^{\circ}$). Hertz and Wood (1988) assume that a $1.4M_\odot$
neutron star is orbiting a low-mass main sequence secondary which
is filling its Roche lobe and use the lack of eclipses to show that the
inclination of GX\,9+9 is less than $\sim 63^{\circ}$. We also note
that the observed semi-amplitude (9.8\%) of the optical modulation
suggests an orbital inclination of less than $70^{\circ}$
\cite{paradijs95}. 
Although the \rxte\
data presented here cannot confirm the X-ray modulation found by Hertz and
Wood (1988), the result is still consistent with the hypothesis that
GX\,9+9 is a low inclination 
system. In order to account for the small amplitude X-ray variability on the
orbital period, \scite{hertz88} suggested that the disc does not hide
the central neutron star (since it is at low inclination angle) and instead
we are actually observing directly the X-ray emission from the vicinity of
the central source. Therefore \scite{hertz88} proposed a variable
covering of an accretion disc corona by the edge of the disc. Alternatively, Schaefer (1990)
suggested that the X-ray modulation is caused by the changing visibility of
X-ray flux reflected off the bright spot although it is more likely
that this mechanism could explain the optical modulation.

\subsection{The Origin of the Optical Modulation}
The observed orbital period of GX\,9+9 provides a valuable tool for understanding the system geometry. Assuming a canonical value for the mass of a neutron star ($M_X=1.4\msun$), then from Kepler's third law, 

\begin{equation}
a=1.471 \rsun(1+q)^{1/3}
\end{equation}

where $a$ is the separation between the two stars, $q$ is the mass ratio, i.e., $M_2/M_X$ and $M_2$ is the mass of the companion star. As the companion fills its Roche lobe, the radius of the companion is given by \cite{paczynski71}

\begin{equation}
R_2=0.462 a\left(\frac{q}{1+q}\right)^{1/3}
\end{equation}

From the main-sequence mass-radius relation \cite{patterson84}

\begin{equation}
R_2=1.0 \rsun \left(\frac{M_2}{\msun} \right)^{0.88}=1.344 \rsun q^{0.88}
\end{equation}

Solving the above three equations simultaneously, we obtain $M_2=0.4 
\msun$, $\rsun=0.45$, $q=0.29$ and $a=1.6 \rsun$. This corresponds to an 
early M dwarf (M0 V--M5 V) with a surface temperature of $\sim$~3400~K 
and an absolute magnitude $M_V\approx10$. From the measured 
$V$ magnitude of 16.79, this gives $M_V=3$ (assuming a distance of 5
kpc and $A_V=0.3$). Hence, the observed 
optical flux is substantially larger than that from the M star itself and 
it is clear that the optical flux must be due to X-ray reprocessing in 
the accretion disc and illuminated face of the companion star. There are 
at least three possible 
origins of the optical flux and its variation associated with the orbital 
period: (i) the accretion disc; (ii) the X-ray heating of the companion 
star which is not in the shadow of the disc; and (iii) the bright spot in 
the edge of the accretion disc.
However, we do not have unambiguous phasing 
information for this system and it is impossible to determine the 
variation of these components with respect to orbital phase. 

\scite{haswell00} proposed that GX\,9+9 is a possible
persistent superhumper based on the fact that $q\lta 0.3$ (based on
its short period and compact object mass of at least $1.4M_\odot$) and also the
similarity of the fractional amplitude of the optical modulation to
that of superhumps seen in SXTs (O'Donoghue \& Charles 1996).  Unlike
in the case of cataclysmic 
variables, in which superhumps are caused by a tidally-driven
modulation of the disc's viscous dissipation which varies on the beat
between the orbital and disc precession period, the optical emission
of LMXBs is dominated by X-ray reprocessing. Hence a
different mechanism is required to account for the superhump period. A
possible solution is to associate the superhumps with varying X-ray
reprocessing by changing disc area (Haswell et al. 2000). In other
words, the disc surface area 
varies on the superhump period, resulting in a modulation in the optical
emission.     
If this is the case, we would expect a superhump period which is
slightly (a few per cent) longer than the orbital period. However, 
with the only significant periodicity being that found in the optical
(here and earlier work), the lack of an X-ray or spectroscopic
modulation still leaves open the possibility that 
 GX\,9+9 is a persistent superhumper as the nature of this variation
 cannot be defined. However, if the X-ray period is true, we can rule
 out the superhump interpretation.

\subsection{Accretion Geometry}
The X-ray spectrum of GX\,9+9 has rarely been studied in detail even
though it belongs to one of the few bright X-ray sources known before
{\it Uhuru} 
(\pcite{bradt68}). In general, previous studies (e.g. \pcite{christian97}; 
\pcite{church00}) and this work suggest that two-component models are
more appropriate in fitting the energy spectrum. Such models consist of a
blackbody component plus a Comptonized component (e.g. CPL, TB, CompST and
CompTT). Similar behaviour is also found in other high-luminosity LMXBs
(\pcite{white88}).

The physical interpretation of the origin of such multi-component models is
still poorly understood although observations have been made for many  LMXBs
(e.g., \pcite{white88}; \pcite{barret00}; \pcite{bloser00a};
\pcite{bloser00b}). It is widely accepted that the hard component (i.e. the
Comptonized component) is due to Comptonization of soft photons in a hot corona, while the source of the soft photons which lead to a Compton
up-scattering and the soft thermal component are still the subject of debate (e.g. \pcite{white88}).
There are several possibilities to explain the origin of the soft thermal
component. The most straightforward solution is that this soft component
originates from an optically thick accretion disc or an optically thick
boundary layer where the accreted matter intersects the neutron star surface.
However, \scite{sunyaev86} showed that the luminosity of the
boundary layer should be at least equal to that of the disc if relativistic
effects are included, whereas we have found that the BB contribution 
ranges only from $\sim10-40$\%. 
In fact, many authors have found similar results for other sources
(\pcite{white88}; \pcite{barret00}; \pcite{bloser00a}; \pcite{bloser00b}).
Therefore \scite{guainazzi98} and \scite{barret00} suggest that the weak
thermal component could be due to an optically thick accretion disc, while
the other Comptonized component arises from an optically thin boundary layer.
However, this picture is not suitable to explain the similarities in spectral
behaviour between neutron stars and black holes since a black hole does not have
a physical surface (see \pcite{barret00}). Moreover, the negative result from the DBB model also
suggests that the soft component might not be from the inner accretion disc,
and such a model is not a good description of the \rxte\ data presented here.
Therefore the Comptonized component is unlikely to be due to the boundary layer and
such a disc corona model can likely be ruled out. 

Among several models mentioned here, the present data are more
consistent with a picture in which the soft component is produced by an
optically {\it thick} boundary layer while the hard component is from an
extended hot corona. As discussed above, such a model cannot explain the small
contribution of the blackbody emission. Hence some mechanism should suppress
the boundary layer emission relative to that of the corona and disc. One
possibility is that the rotation period of the neutron star is close to 
equilibrium with the Keplerian period of the inner edge of the accretion disc
(\pcite{white88}) such that little energy is given up in the boundary layer.
Another possibility is that part of the boundary layer emission is blocked by
the inner disc corona or thickened disc (\pcite{lamb86}; \pcite{klis87}).
Such mechanisms have been discussed for the X-ray spectrum of 4U\,1916--05
(\pcite{church98}; \pcite{bloser00a}). In particular, \scite{bloser00a}
have shown that by fitting a BB+CPL model, the blackbody radius of
4U\,1916--05 decreases slightly as the accretion rate increases. In
Fig.~{\ref{fig:spec}, all four models show a decreasing trend for $R_{bb}$ 
as the total flux increases. This may indicate that the inner disc can become
thicker as the accretion increases and thereby obscure a larger region of 
the
boundary layer. In addition, the derived $R_{bb}$ here is consistent with
$\sim$~2--3~km, implying that the blackbody component can come from only a
limited region of the neutron star surface such as an optically thick
boundary layer. It is also interesting to note that GX\,9+1
\cite{white86}, 4U\,1735--444 \cite{seon97} and 4U\,1820--30 
\cite{bloser00b}, which are all luminous atoll sources, show similar
blackbody radii in the persistent spectrum.  Spectral analysis of these
sources also suggests that the blackbody component comes from an optically thick boundary layer.

Applying the above accretion geometry to describe GX\,9+9, one would
expect to detect an iron $K$ emission line at 6.4--6.7~keV which is a common
feature of the spectra of many luminous LMXBs (e.g. \pcite{white86};
\pcite{barret00}), as a result of the accretion disc corona (see
\pcite{fabian82}). However, observations have shown  that such emission is 
not present in GX\,9+9 (see  \pcite{church00}).
Interestingly, GX\,9+1 is  also well-known for its lack of iron
$K$ emission \cite{white86}. A reflection model ({\it pexrav} in {\tt XSPEC}) was
also used to fit the spectra so as to search for a reflection component from the disc,
but the derived reflection factor was consistent with zero and thus there is
no evidence of reflection. One possible reason for this is the relatively low
corona temperature such that the hard photons from the corona will tend to be
photoelectrically absorbed by the matter in the disc, rather than Compton
scattered. The lack of the line emission and reflection component indicates
that a significant fraction of the disc cannot be seen, either by
reflection or irradiation. 

Regarding a physical interpretation of the spectral models, a thermal bremsstrahlung
interpretation for the hard component seems to be unrealistic since the
derived radius of the plasma cloud is too far from the neutron star ($\sim
10^9$~cm) to be heated to several keV by the neutron star (\pcite{white86}).
On the other hand, although the CompTT model is an improved version of CompST
which includes relativistic effects and extends to the cases of both
optically thin and thick plasmas, the larger error bars of optical
depth and electron temperature compared to CompST
make the CompST model more suitable to describe the present \rxte\ data. 
Therefore, we will refer to the CompST model when discussing Comptonization below.  For the CPL model, it approximates an
unsaturated Comptonized spectrum ($y=4kT_e \tau^2/m_e c^2\approx 1$) of cool
photons up-scattered in a hot electron plasma and its results may not be
directly interpretable in terms of a physical description. Therefore the
CompST model can give a more complete description of the source spectrum. In
fact, several luminous X-ray sources (such as 4U\,1636--536, 4U\,1820--30,
Sco\,X--1, Sco\,X--2, Cyg\,X--2, GX\,9+1 and GX\,17+2) have shown that
the CompST and CPL models provide equally good fits to the source spectrum
(\pcite{christian97}). It is also interesting to note that the derived $kT_e$
and $\tau$ for Sco\,X--1, 4U\,1735-444, 4U\,1916--05, 4U\,1820--30,
Cyg\,X--2, GX\,9+1, GX\,17+2 and KS\,1731--260 are very similar to the
results presented here (see Table~\ref{tab:9+9compare}).

\begin{table*}
\caption{BB+CompST spectral fits of other luminous LMXBs.}
\label{tab:9+9compare}
\begin{center}
\begin{tabular}{c c c c l}
\hline
Source & $kT_{bb}$ & $kT_e$ & $\tau$ & Reference\\
       &  (keV)    &  (keV) &        &          \\
\hline
Sco\,X--1& 1.6 & 3 & 10.5 & \pcite{white85}\\
4U\,1735--444& 1.45--1.90 & 3.05--3.26 & 11.93--13.99 & \pcite{seon97}\\
4U\,1916--05$^a$ & 1.49--2.03 & 3.53--8.85 & 5.78--13.46 & \pcite{bloser00a}\\
4U\,1820--30$^a$ & 1.30--2.36 & 2.75--7.94 & 6.00--15.21 & \pcite{bloser00b}\\
Cyg\,X--2 & 1.13 & 3.75 & 8.4 & \pcite{christian97}\\
GX\,9+1 & 1.49 & 3.00 &  10.9 & \pcite{white88}\\
GX\,17+2 & 1.18 & 3.00 & 13.0 & \pcite{white88}\\
KS\,1731--260 & 1.00--1.30 & 2.60--2.80 & 10.40--12.60 & \pcite{barret00}\\
GX\,9+9 & 1.37--1.96 & 2.81--3.38 & 9.12--12.18 & this work\\
\hline
\\
\multicolumn{5}{l}{$^a$ Use CompTT instead of CompST for the Comptonized component}
\end{tabular}
\end{center}
\end{table*}

The high-quality \rxte\ data presented here can be summarised with the
following physical model: in the upper banana state, the luminosity is
relatively high and the blackbody radiation comes from an optically thick
boundary layer which is partly obscured by the inner accretion disc,
producing $\sim$~10--40\% blackbody flux from a small equatorial region of
the neutron star. 
Increases in the blackbody flux are also responsible for
the flaring activity. 
Similar behaviour is also found in Sco\,X--1
\cite{white85} and Sco\,X--2 \cite{white86}. A hot electron cloud
($T\approx 3.5\times 10^7$~K) in the form of an accretion disc corona
surrounds the neutron star and at such high luminosity, the blackbody photons
from the boundary layer can heat up the corona to maintain its temperature
(see Fig.~\ref{fig:spec}) at about a constant level in spite of the
Compton cooling. The relatively strong blackbody radiation produces more extensive
ionisation and hence increases the optical depth $\tau$ of the corona, 
reducing the radiation drag that generates QPOs in the framework of
the sonic-point model (\pcite{miller98}), and thereby accounting for
the lack of kHz QPO in GX\,9+9. It is also known that when other atoll sources (e.g. 4U\,1636--53: \pcite{wijnands97a}; 4U\,1820--30: \pcite{smale97}) are in their upper banana branch, kHz QPOs do not occur. 

\section{Conclusions}

Simultaneous \rxte\ and optical high-speed observations were performed to
study the atoll source GX\,9+9. The high-quality optical light curves from
high-speed CCD photometry during a four-night observation confirm, but
with much greater precision, the 4.19-hr 
orbital period \cite{schaefer90}. However, our \rxte\ observations cannot
confirm the X-ray modulation at the orbital period \cite{hertz88}. By
using archival \asca\ and \exosat\ data, and also our recent \sax\
observation, no X-ray modulation could be 
revealed. GX\,9+9 also shows variability on  timescales 
$\sim$~1.4~hr in the optical light curves which could be related to the  
mHz QPO as
seen in other LMXBs. We do not find any significant X-ray/optical correlation in the light curves.
Observations indicate that the inclination angle
of GX\,9+9 
cannot be too high ($<$~70$^{\circ}$). The observed optical flux is mainly 
due to X-ray
irradiation of the outer accretion disc and the reprocessing of a
bright spot in the disc and X-ray heating of the companion can
contribute the observed modulation.   
The CD and HID show that GX\,9+9
was in the upper banana state during observations. Time-resolved X-ray
spectroscopy was also performed and several two-component spectral models (BB+CPL,
BB+TB, BB+CompST and CompTT) can provide a good fit to the energy spectra.
A two-component spectral model is always required to fit the X-ray emission from
several luminous LMXBs (see Table~\ref{tab:9+9compare}). In general, such models
consist of a blackbody component (soft component) and a Comptonized component
(hard component). The blackbody component is often interpreted as the
contribution from either an optically thick accretion disc or an optically
thick boundary layer (interaction between the inner accretion disc and the
neutron star's surface). On the other hand, the hard component may be from a
hot extended corona, optically thin boundary layer.
The likely interpretation for GX\,9+9 can be understood in
terms of an optically thick boundary layer (soft component) which is partly
obscured by the inner disc or thickened disc, while the hard Comptonized
component is from an extended hot corona.

\section*{Acknowledgments}
We thank Christine Done for the red-noise generator code, and the
\rxte\ SOC team for their efforts in schuduling the simultaneous
observations.

\end{document}